\documentclass[english,aps, reprint, twocolumn]{revtex4}
\usepackage[T1]{fontenc}
\usepackage[latin9]{luainputenc}
\usepackage{bm}
\usepackage{amsmath}
\usepackage{amssymb}
\usepackage{graphicx}
\usepackage{placeins}

\makeatletter
\@ifundefined{textcolor}{}
{%
 \definecolor{BLACK}{gray}{0}
 \definecolor{WHITE}{gray}{1}
 \definecolor{RED}{rgb}{1,0,0}
 \definecolor{GREEN}{rgb}{0,1,0}
 \definecolor{BLUE}{rgb}{0,0,1}
 \definecolor{CYAN}{cmyk}{1,0,0,0}
 \definecolor{MAGENTA}{cmyk}{0,1,0,0}
 \definecolor{YELLOW}{cmyk}{0,0,1,0}
}

\makeatother

\usepackage{babel}
\begin{document}

\preprint{This line only printed with preprint option}

\title{Magnon heralding in cavity optomagnonics}

\author{Victor A. S. V. Bittencourt}
\email{victor.bittencourt@mpl.mpg.de}

\affiliation{Max Planck Institute for the Science of Light, Staudtstr. 2, PLZ
91058, Erlangen, Germany}

\author{Verena Feulner}

\affiliation{Max Planck Institute for the Science of Light, Staudtstr. 2, PLZ
91058, Erlangen, Germany}

\author{Silvia Viola Kusminskiy}
\email{silvia.viola-kusminskiy@mpl.mpg.de}

\affiliation{Max Planck Institute for the Science of Light, Staudtstr. 2, PLZ
91058, Erlangen, Germany}
\affiliation{Institute for Theoretical Physics, University Erlangen-Nürnberg, Staudtstr. 7, 91058 Erlangen, Germany}
\begin{abstract}
In the emerging field of cavity optomagnonics, photons are coupled coherently to magnons in solid-state systems. These new systems are promising for implementing hybrid quantum technologies. Being able to prepare Fock states in such platforms is an essential step towards the implementation of quantum information schemes. We propose a magnon-heralding protocol to generate a magnon Fock state by detecting an optical cavity photon. Due to the peculiarities of the optomagnonic coupling, the protocol involves two distinct cavity photon modes. Solving the quantum Langevin equations of the coupled system, we show that the temporal scale of the heralding is governed by the magnon-photon cooperativity and derive the requirements for generating high fidelity magnon Fock states. We show that the nonclassical character of the heralded state, which is imprinted in the autocorrelation of an optical ``read'' mode, is only limited by the magnon lifetime for small enough temperatures. We address the detrimental effects of nonvacuum initial states, showing that high fidelity Fock states can be achieved by actively cooling the system prior to the protocol.
\end{abstract}
\maketitle

\section{introduction}

Hybrid systems play an important role in the ongoing development of quantum technologies, for example as interfaces between different types of information carriers and between storage and transmission lines \citep{kurizkiQuantumTechnologiesHybrid2015a}. A new exciting development in this area is the recently demonstrated possibility of coherently coupling photons to collective magnetic excitations (magnons) in magnetically ordered solid-state systems, both for microwave \citep{bourhillUltrahighCooperativityInteractions2016,tabuchiHybridizingFerromagneticMagnons2014,zhangStronglyCoupledMagnons2014} and optical photons \citep{haighMagnetoopticalCouplingWhisperinggallerymode2015,zhangOptomagnonicWhisperingGallery2016}. In these systems, the spin-photon coupling is enhanced due to the collective character of the magnetic excitations, as well as by the usage of a cavity for the photons. Applying an external magnetic field allows one moreover to tune the frequency of the magnonic excitations. This has been used to bring magnon modes in resonance with photons in a microwave cavity, which allowed for the observation of strong coupling between magnons and photons \citep{bourhillUltrahighCooperativityInteractions2016,tabuchiHybridizingFerromagneticMagnons2014,zhangStronglyCoupledMagnons2014,goryachevHighCooperativityCavityQED2014, Huebl2013}. In turn, this coupling has been used for engineering the indirect interaction between the magnons and a superconducting qubit \citep{lachance-quirionResolvingQuantaCollective2017b,tabuchiCoherentCouplingFerromagnetic2015}. The coherent interaction between solid-state magnons and optical photons has been observed recently in Brillouin light scattering experiments in yttrium-iron-garnet (YIG) optical cavities \citep{haighMagnetoopticalCouplingWhisperinggallerymode2015,osadaCavityOptomagnonicsSpinOrbit2016,zhangOptomagnonicWhisperingGallery2016,osadaBrillouinLightScattering2018,haighSelectionRulesCavityenhanced2018,haighTripleResonantBrillouinLight2016}, and theoretically studied \citep{liuOptomagnonicsMagneticSolids2016,violakusminskiyCoupledSpinlightDynamics2016,sharmaLightScatteringMagnons2017,osadaOrbitalAngularMomentum2018,grafCavityOptomagnonicsMagnetic2018a}. In this framework, the solid-state system is both the host of the magnetic excitations and the cavity supporting the photons. The origin of the optomagnonic coupling is the Faraday effect, in which the light which propagates in a magnetized material has its polarization rotated \citep{landauElectrodynamicsContinuousMedia1984}. In contrast to the microwave regime, optical photons and magnons couple parametrically \citep{liuOptomagnonicsMagneticSolids2016,violakusminskiyCoupledSpinlightDynamics2016}.

\begin{figure}
\centering{}\includegraphics[width=1\columnwidth]{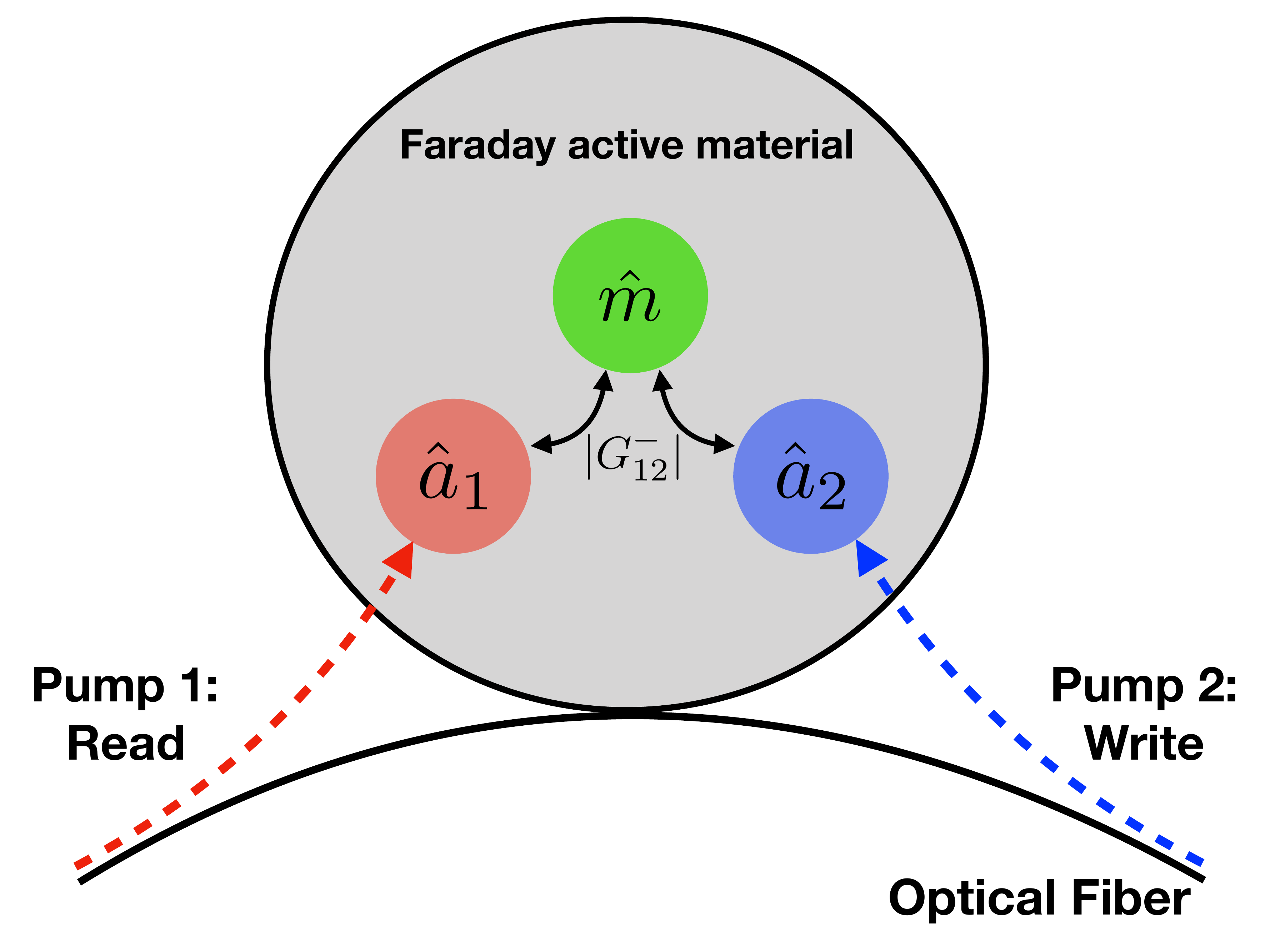}
\caption{Proposed setup for the magnon heralding protocol. Two optical modes $\hat{a}_{1}$ and $\hat{a}_{2}$ are coupled to a magnon mode $\hat{m}$ via the optomagnonic interaction $G_{12}^{-}\hat{a}_{1}^{\dagger}\hat{a}_{2}\hat{m}^{\dagger}+h.c.$. Each mode can be individually driven externally, in order to ``write'' a magnon Fock state by measuring one photon and subsequently ``read'' it.}
\label{Scheme}
\end{figure}
An important part in engineering quantum devices is state preparation. In this manuscript, we propose a heralding protocol in a cavity optomagnonic system in which a magnon Fock state is created by the measurement of an optical photon. Apart from their interest for quantum information processing \citep{andersenHybridDiscreteContinuousvariable2015}, magnon Fock states are collective excitations involving millions of spins, and the heralding protocol can be used as a preparation step to probe quantum mechanics in macroscopic systems. Heralding protocols are often proposed in the context of hybrid systems coupled to light, inspired by the DLCZ protocol \citep{duanLongdistanceQuantumCommunication2001}. The method relies on generating an entangled state of a given system of interest and light, which is projected to a desired configuration once a photon is measured. Heralding was originally proposed to generate entanglement between collective excitations in atomic clouds and was further translated to other systems, for example for preparing single phonon Fock states in optomechanics \citep{gallandHeraldedSinglePhononPreparation2014,hongHanburyBrownTwiss2017,riedingerNonclassicalCorrelationsSingle2016}, or to prepare atomic states in cold-atoms experiments \citep{wolfgrammAtomResonantHeraldedSingle2011,welteCavityCarvingAtomic2017,mcconnellGeneratingEntangledSpin2013,mcconnellEntanglementNegativeWigner2015,chuuQuantumMemoryOptically2008}. In our case, due to the peculiarities of the optomagnonic coupling, the heralding protocol involves two photon modes and one magnon mode, as depicted schematically in Fig.~\ref{Scheme}. The protocol consists of two phases: write and read, each implemented with light pulses. In the write phase, correlated pairs of magnons and photons are created, and the measurement of a photon collapses the state of the system to a magnon Fock state. The read phase maps the heralded excitation to a photon which can be further probed or used for other purposes. 

We present an analytical analysis of the proposed protocol and study its feasibility in cavity optomagnonic solid state systems. We show that the probability of heralding a magnon Fock state can be in line with the experimentally measured heralding probability in optomechanical experiments \citep{riedingerNonclassicalCorrelationsSingle2016,riedingerRemoteQuantumEntanglement2018} and in cold atoms experiments \citep{christensenQuantumInterferenceSingle2014,matsukevichDeterministicSinglePhotons2006,mcconnellEntanglementNegativeWigner2015}, provided cooperativities of the order of $10^{-2}$ can be achieved. The read photon field is a witness of heralding, exhibiting non classical counting statistics for a successfully heralded magnon state. If the strong-coupling regime is reached, Rabi oscillations take place, allowing for an efficient conversion between the heralded magnon state and the read photon field. We moreover study the dependence of the protocol on the initial state of the system and derive cooling requirements. Our results show that, although the heralding protocol is highly susceptible to deviations of the initial state from the magnon vacuum, a high fidelity heralded single-magnon Fock state can be achieved through an efficient initial cooling of a thermal magnon state. We complement our analytical results based on square light pulses, with numerical results for Gaussian pulses.

The manuscript is organized as follows. In Sec. II we present the model based on a linearized optomagnonic Hamiltonian. In Sec. III we describe the heralding protocol and show the temporal constraints imposed by both magnon and photon linewidths. Sec. IV is devoted to the analytical analysis of the dynamics of the system based on the linear quantum Langevin equations. In Sec. V we present the results and analyze the impact of the initial state on the protocol, deriving the cooling requirements. Finally we present our conclusions and future perspectives. Details of the calculations and numerical results for the protocol involving Gaussian beams are presented in the Appendixes.

\section{Model}

The coupling between light and magnetization in a Faraday-active material manifests itself in a modification of the electromagnetic energy by the term \citep{landauElectrodynamicsContinuousMedia1984}
\begin{equation}
\bar{U}=-i\frac{\theta_{F}\lambda_{n}}{4\pi}\varepsilon_{0}\varepsilon\int d\bm{r}\bm{M}(\bm{r},t)\cdot[\bm{E^{*}}(\bm{r},t)\times\bm{E}(\bm{r},t)]\,,\label{eq:OMEnergy}
\end{equation}
where $\bm{M}(\bm{r},t)$ is the magnetization of the material in units of the saturation magnetization, and the complex representation of the electric field is used. The material-specific constant $\theta_{F}\lambda_{n}/2\,\pi$ is given in terms of the Faraday rotation angle $\theta_{F}$ per wavelength $\lambda_{n}$ in the material with relative permittivity $\varepsilon$, and $\varepsilon_{0}$ is the vacuum permittivity. The optomagnonic coupling can also have a contribution from the Cotton-Mouton effect (or magnetic linear birefringence) \cite{landauElectrodynamicsContinuousMedia1984,liuOptomagnonicsMagneticSolids2016}. For simplicity we do not consider this contribution, since its main effect consists on a renormalization of the coupling constants and does not affect our results.

The optomagnonic Hamiltonian, describing the interaction between magnons and optical photons, is obtained by quantizing Eq.~\eqref{eq:OMEnergy} \citep{violakusminskiyCoupledSpinlightDynamics2016,grafCavityOptomagnonicsMagnetic2018a}. For this purpose we consider that the material acts as an optical cavity and we quantize the electromagnetic field in terms of creation an annihilation operators of its eigenmodes $\bm{E}(\bm{r},t)\rightarrow\hat{\bm{E}}^{(+)}(\bm{r},t)=\sum_{i}\bm{E}_{i}(\bm{r})\hat{a}_{i}(t)$ and $\bm{E}^{*}(\bm{r},t)\rightarrow\hat{\bm{E}}^{(-)}(\bm{r},t)=\sum_{i}\bm{E}_{i}^{*}(\bm{r})\hat{a}_{i}^{\dagger}(t)$. The spin wave part of the magnetization $\bm{M}(\bm{r},t)$ is described by fluctuations $\bm{m}(\bm{r},t)$ on top of a ground state $\bm{m}_{S}(\bm{r})$,
\begin{equation}
\bm{M}(\bm{r},t)=\bm{m}_{S}(\bm{r})+\bm{m}(\bm{r},t).\label{eq:Mm}
\end{equation}
In the limit of small deviations $\vert\bm{m}\vert\ll1$, we can treat the fluctuations as harmonic oscillators and quantize the field $\bm{m}(\bm{r},t)\rightarrow\hat{\bm{m}}(\bm{r},t)$ akin to the quantization of lattice vibrations (phonons) \citep{millsQuantumTheorySpin2006}
\begin{equation}
\hat{\bm{\bm{m}}}(\bm{r},t)={\displaystyle \sum_{k}\left[\bm{m}_{k}(\bm{r})\:\hat{m}_{k}\,e^{-i\,\Omega_{k}\:t}+\bm{m}_{k}^{*}(\bm{r})\,\hat{m}_{k}^{\dagger}\,e^{i\,\Omega_{k}\,t}\right]\,,}\label{eq:quant_m}
\end{equation}
where $k$ labels the magnon modes with frequency $\Omega_{k}$ and the creation and annihilation operators satisfy the bosonic commutation relations $[\hat{m}_{k},\hat{m}_{k^{\prime}}^{\dagger}]=\delta_{k\,k^{\prime}}$ and
$[\hat{m}_{k},\hat{m}_{k^{\prime}}]=[\hat{m}_{k}^{\dagger},\hat{m}_{k^{\prime}}^{\dagger}]=0$. This quantization procedure gives the \emph{optomagnonic Hamiltonian}
\begin{eqnarray}
\hat{H} & = & {\displaystyle \sum_{i}}\hbar\omega_{i}\hat{a}_{i}^{\dagger}\hat{a}_{i}+{\displaystyle \sum_{k}}\hbar\Omega_{k}\hat{m}_{k}^{\dagger}\hat{m}_{k} \nonumber \\
 & + & {\displaystyle \hbar\sum_{i,j,k}}\hat{a}_{i}^{\dagger}\hat{a}_j(G_{ijk}^{+}\hat{m}_{k}+G_{ijk}^{-}\hat{m}_{k}^{\dagger}), \label{eq:OMHamiltonian-1}
\end{eqnarray}
with the first and second terms corresponding to the non-interacting part of, respectively, the photon and magnon field dynamics. The couplings $G_{ijk}^{+}=(G_{jik}^{-})^{*}$ are given by \citep{grafCavityOptomagnonicsMagnetic2018a}
\begin{eqnarray}
G_{ijk}^{+} & = & \frac{\theta_{F}\lambda_{n}}{4\pi i\hbar}\varepsilon_{0}\varepsilon{\displaystyle \int_{V}}{\rm d}^{3}\bm{r}\bm{m}_{k}(\bm{r})\cdot\bm{E}_{i}^{*}(\bm{r})\times\bm{E}_{j}(\bm{r})\,.\label{eq:CouplingIntegral}
\end{eqnarray}
The interacting part of Eq.~\eqref{eq:OMHamiltonian-1} describes a process in which one photon in a mode $j$ is annihilated creating a photon in the mode $i$ and a magnon in the mode $k$, and the complementary process in which a magnon $k$ and a photon $j$ are annihilated creating a photon $i$.

In the following we will consider the case in which two non-degenerate photon modes interact with one magnon mode; see Fig.~\ref{Scheme}. This is valid, for example, for recent experiments with YIG spheres \citep{haighSelectionRulesCavityenhanced2018,haighTripleResonantBrillouinLight2016,osadaBrillouinLightScattering2018,osadaCavityOptomagnonicsSpinOrbit2016,osadaOrbitalAngularMomentum2018,sharmaLightScatteringMagnons2017}. In this case, the photon modes would correspond to counter propagating whispering gallery modes (WGM) of the optical field. These possess different polarizations due to spin-orbit coupling and are nondegenerate due to geometric birefringence. We describe the coupled magnon-photon dynamics via a linearized optomagnonic Hamiltonian. Note that the linearization is twofold: we consider the fluctuations of the photon fields around their steady-state values, and of the magnon field around a magnetic ground state as given by Eqs. \eqref{eq:Mm} and \eqref{eq:quant_m} and already used to write Eq.~\eqref{eq:OMHamiltonian-1}. The linearized Hamiltonian in the resolved sideband regime is given by (see Appendix A)
\begin{align}
\hat{H}_{{\rm tot}} & =\hbar\alpha_{1}^{*}\hat{a}_{2}\left[G_{12}^{+}\hat{m}e^{i(\Delta_{2}-\Omega)t}+G_{12}^{-}\hat{m}^{\dagger}e^{i(\Delta_{2}+\Omega)t}\right]\nonumber \\
 & +\hbar\alpha_{2}\hat{a}_{1}^{\dagger}\left[G_{12}^{+}\hat{m}e^{-i(\Delta_{1}+\Omega)t}+G_{12}^{-}\hat{m}^{\dagger}e^{-i(\Delta_{1}-\Omega)t}\right]\nonumber \\
 & +h.c.,\label{eq:TimeDependentHamiltonian-1-1}
\end{align}
where we have labeled the photon modes as 1 and 2, $\Delta_{i}=\omega_{L}-\omega_{i}$ ($i=1,2$) are the detunings between the laser frequency and the respective mode frequency, and $\alpha_{i}=-\frac{\epsilon_{i}}{i\Delta_{i}-\frac{\kappa_{i}}{2}}$, with $\epsilon_{i}=\hbar\sqrt{\frac{2\kappa_{i}\mathcal{P}_{i}}{\hbar\omega_{L}}}$ depending on the driving laser power $\mathcal{P}_{i}$ and on the coupling between the pumped mode and the fiber $\kappa_{i}$. At resonance, $\Delta_{i}=0$ and $\alpha_{i}$ is directly related to the laser power through $\epsilon_{i}$ and, consequently, to the average number of photons inside the cavity.

\begin{figure}
\centering{}\includegraphics[width=1\columnwidth]{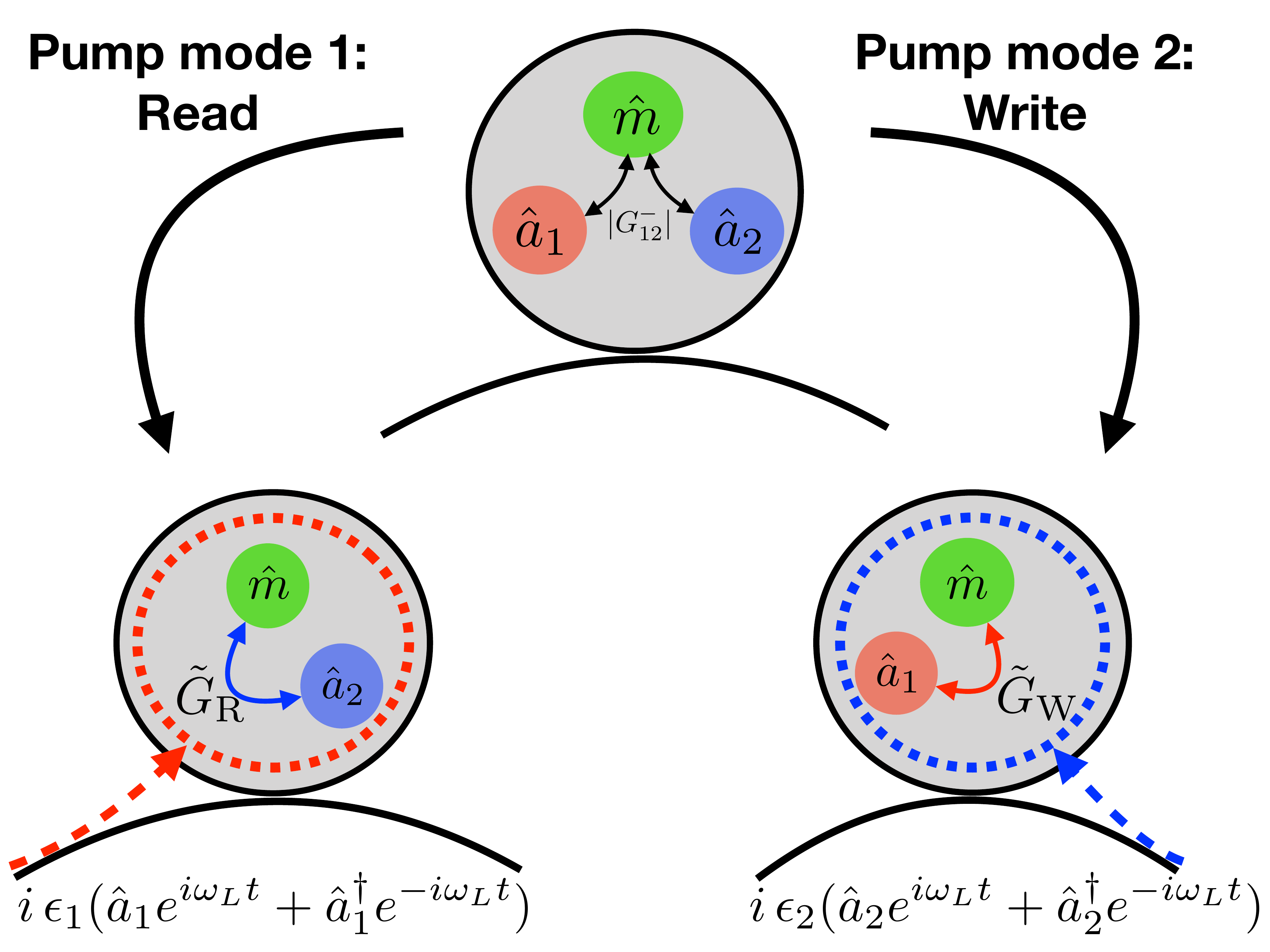}
\caption{Illustration of the different interactions used for heralding. When mode $2$ is driven at resonance $\omega_{{\rm L}}\sim\omega_{2}=\omega_{1}+\Omega$, $\alpha_{1}=0$, and mode $1$ couples with the magnon mode with strength $\tilde{G}_{{\rm W}}=\alpha_{2}^{*}G_{21}^{+}$, where $\alpha_{2}$ depends on the power of the pumping laser. Instead, if mode $1$ is pumped then $\omega_{{\rm L}}\sim\omega_{1}=\omega_{2}-\Omega$, $\alpha_{2}=0$, and the interaction between mode $2$ and the magnon mode is driven with an enhanced coupling $\tilde{G}_{{\rm R}}=\alpha_{1}^{*}G_{12}^{-}$.}
\label{Linearization}
\end{figure}

The optomagnonic coupling in solid state systems is subject to the usual energy conservation requirements, $\omega_{i}=\omega_{j}\pm\Omega_{k}$, and also to selection rules involving conservation of angular momentum \citep{haighSelectionRulesCavityenhanced2018,haighTripleResonantBrillouinLight2016,sharmaLightScatteringMagnons2017,osadaCavityOptomagnonicsSpinOrbit2016,osadaBrillouinLightScattering2018,osadaOrbitalAngularMomentum2018}. We consider that the selection rules manifest themselves as a coupling asymmetry $G_{12}^{-}\gg G_{12}^{+}\sim0$, meaning that the creation and annihilation processes are unbalanced with respect to the modes involved. This nonreciprocity between processes involving different polarizations has been observed in YIG spheres, reflected in an asymmetry in the Stokes and anti-Stokes lines in Brillouin light scattering experiments \citep{osadaOrbitalAngularMomentum2018,osadaCavityOptomagnonicsSpinOrbit2016,osadaBrillouinLightScattering2018,haighSelectionRulesCavityenhanced2018}. In a setup in which light propagates perpendicularly to the saturation magnetization, the non reciprocity in the couplings can be further modified by the inclusion of the Cotton-Mouton effect terms, which does not modify the selection rules \citep{sharmaLightScatteringMagnons2017}. Due to this asymmetry, the only two possible processes are a creation of a photon in mode $2$ through the annihilation of a photon in mode $1$ and a magnon, and the complementary process. The optomagnonic Hamiltonian Eq.~\eqref{eq:TimeDependentHamiltonian-1-1} then reads
\begin{equation}
\hat{H}_{{\rm tot}}=\hbar G_{12}^{-}m^{\dagger}e^{i\Omega t}(\alpha_{1}^{*}\hat{a}_{2}e^{i\Delta_{2}t}+\alpha_{2}\hat{a}_{1}^{\dagger}e^{-i\Delta_{1}t})+h.c.,\label{eq:AsHam}
\end{equation}
which contains two resonances: (i) $\Delta_{1}=\Omega$, driven by pumping mode $2$, and (ii) $\Delta_{2}=-\Omega$, driven by pumping mode $1$. Those resonances are schematically depicted in Fig.~\ref{Linearization} [note that $\left(G_{12}^{-}\right)^{*}=G_{21}^{+}$ ]. The write and read phases of the heralding protocol are thus implemented by driving these interactions.

\section{Protocol}

\begin{figure}
\centering{}\includegraphics[width=1\columnwidth]{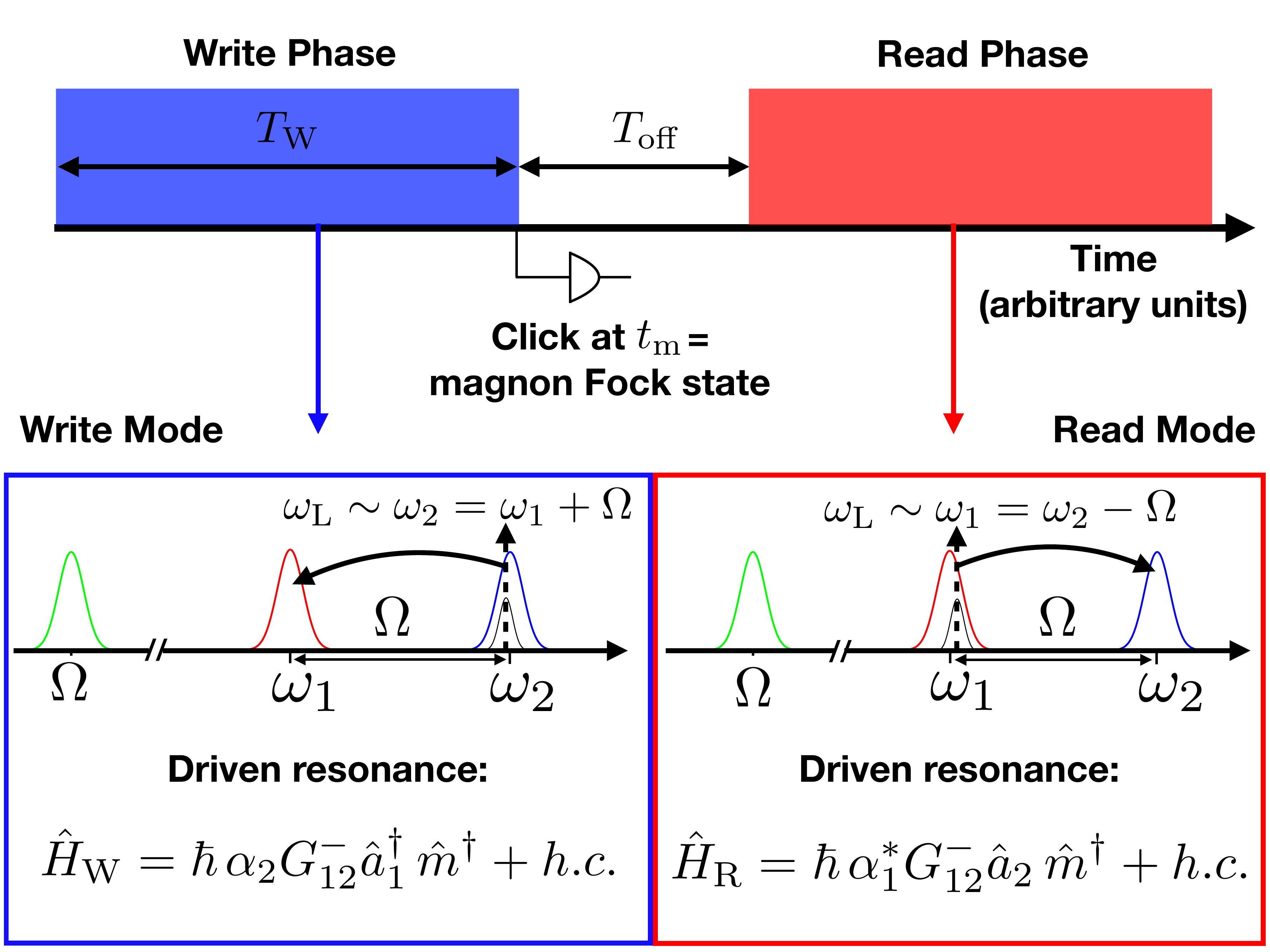}
\caption{Depiction of the heralding protocol and the write and read modes scheme. The write mode is turned on by pumping mode $2$ at resonance for a period $T_{{\rm W}}$. A single magnon Fock state is post-selected by the measurement of a write photon at $t_{{\rm m}}$. After an off period $T_{{\rm off}}$ the read mode is turned on by pumping mode $1$ at resonance.}
\label{Resonances}
\end{figure}

We now proceed to detail the heralded write-and-read protocol, depicted schematically in Fig.~\ref{Resonances}. Starting from Eq.~\eqref{eq:AsHam}, by pumping the optical mode $2$ at resonance, $\omega_{L}\sim\omega_{2}=\omega_{1}+\Omega$ and therefore $\Delta_{1}=\Omega$. Since mode 1 is not driven, $\alpha_{1}=0$ and the Hamiltonian from Eq.~\eqref{eq:AsHam} reads
\begin{equation}
\hat{H}\rightarrow\hat{H}_{{\rm W}}=\hbar(\alpha_{2}^{*}G_{21}^{+}\hat{a}_{1}\hat{m}+\alpha_{2}G_{12}^{-}\hat{a}_{1}^{\dagger}\hat{m}^{\dagger}),\label{eq:PA01}
\end{equation}
which is a two-mode parametric amplifier between the magnon mode and the cavity mode $1$. If the evolution under Eq.~\eqref{eq:PA01} takes place for a period $T\ll\vert\alpha_{2}G_{12}^{-}\vert^{-1}$, and disregarding thermal effects for now, an initial ground state $\vert\psi_{0}\rangle=\vert0\rangle_{a_{1}}\otimes\vert0\rangle_{a_{2}}\otimes\vert0\rangle_{m}$ evolves to
\begin{eqnarray*}
\vert\psi_{W}(T)\rangle & \simeq & \frac{\vert0\rangle_{a_{1}}\vert0\rangle_{a_{2}}\vert0\rangle_{m}-(i\alpha_{2}G_{12}^{-}T)\vert1\rangle_{a_{1}}\vert0\rangle_{a_{2}}\vert1\rangle_{m}}{\sqrt{1+\vert\alpha_{2}G_{12}^{-}\vert^{2}T^{2}}}
\end{eqnarray*}
and \textbf{$\vert\alpha_{2}G_{12}^{-}\vert^{2}T^{2}/\left(1+\vert\alpha_{2}G_{12}^{-}\vert^{2}T^{2}\right) = p_{1}$} is the probability for a pair of excitations to be created by $\hat{H}_{{\rm W}}$. A projective measurement of a photon in mode 1 collapses the state to a single-magnon state with small probability $p_{1}$. We refer therefore to $\hat{H}_{{\rm W}}$ as the ``write'' Hamiltonian.

We can also turn our system into a ``reading mode'' by driving instead the optical mode $1$ with $\omega_{L}\sim\omega_{1}=\omega_{2}-\Omega$. In this case, $\Delta_{2}=-\Omega$ and the driven resonance of Eq.~\eqref{eq:AsHam} is 
\begin{equation}
\hat{H}\rightarrow\hat{H}_{{\rm R}}=\hbar(\alpha_{1}^{*}G_{12}^{-}\hat{a}_{2}\hat{m}^{\dagger}+\alpha_{1}G_{21}^{+}\hat{a}_{2}^{\dagger}\hat{m}),\label{eq:BS01}
\end{equation}
a beam-splitter interaction between the magnon mode and the cavity mode $2$. Such dynamics drives magnon-photon oscillations with frequency $\vert\alpha_{1}G_{12}^{-}\vert$, mapping the excitation in the magnon mode to a photon excitation in mode 2. In the weak-coupling regime $\vert\alpha_{1}G_{12}^{-}\vert$ is smaller than the cavity linewidth $\kappa$, and the oscillations are suppressed by the photon decay. Otherwise, in the strong coupling regime $\vert\alpha_{1}G_{12}^{-}\vert > \kappa$ the magnon-photon oscillations allow the read out of the heralded state. Considering Eqs. \eqref{eq:PA01} and \eqref{eq:BS01}, the magnon heralding protocol is implemented as follows: first the system is prepared near its ground state. By driving mode $2$ at resonance, the parametric amplifier Hamiltonian Eq.~\eqref{eq:PA01} is tuned, generating correlated pairs of write mode photons and magnons. For weak coupling, the measurement of a single photon will collapse the system to a single magnon state with probability $p_{1}$. After an interval without driving, the read Hamiltonian Eq.~\eqref{eq:BS01} is tuned by driving the mode $1$ at resonance, transferring magnons to read-mode photons. This reading step requires stronger coupling between the magnon and the photon mode, which can be achieved by increasing the pumping laser power encoded in $\alpha_{1}$, and which is limited by the number of photons supported by the cavity. The read-photon state can be probed via interferometric techniques, certifying the nonclassicality of the heralded state. The protocol is depicted in Fig.~\ref{Resonances} with the frequency scheme of the write and read modes needed for the implementation. We emphasize that, since the considered optomagnonic coupling connects two distinct optical modes, the interaction between the magnon mode and a given optical mode is always driven by pumping the other optical mode. 

To end this section we comment briefly on the protocol requirements. Since the laser pulse is limited by the cavity linewidth, the spectral width of the write pulse needs to be narrower than the cavity linewidth. Moreover, assuming that the magnon mode with a linewidth $\gamma$ interacts with a bath characterized by a mean number of excitations $n_{{\rm Th}}$, to avoid thermalization of the heralded magnon state one must impose an interval between write and read pulses $T_{{\rm off}}<1/n_{{\rm Th}}\gamma$. Therefore, the total time between the beginning of the protocol and the start of the read pulse $T_{{\rm W}}+T_{\rm{off}}$ needs to satisfy \citep{gallandHeraldedSinglePhononPreparation2014}
\begin{equation}
1/\kappa < T_{{\rm W}}+T_{\rm{off} }< 1/n_{{\rm Th}}\gamma.
\label{eq:TimeConstraints}
\end{equation}

The above protocol is akin to the one proposed and implemented in optomechanical systems to generate single-phonon states and to herald entanglement \citep{gallandHeraldedSinglePhononPreparation2014,hongHanburyBrownTwiss2017,riedingerNonclassicalCorrelationsSingle2016}, and to the one used in cold atoms systems \citep{chuuQuantumMemoryOptically2008,mcconnellEntanglementNegativeWigner2015,mcconnellGeneratingEntangledSpin2013}. The necessity of using two different photon modes is a characteristic of the optomagnonic system, and could be used in a similar fashion to the polarization dependent transition of cold atoms to engineer nonreciprocal devices \citep{jungeStrongCouplingSingle2013,petersenChiralNanophotonicWaveguide2014,sayrinNanophotonicOpticalIsolator2015,scheucherQuantumOpticalCirculator2016}.

\section{Analytical analysis}

The dynamics of the creation and annihilation operators under the write and read dynamics is described through linear quantum Langevin equations (QLE). The QLE under the dynamics of the write Hamiltonian Eq.~\eqref{eq:PA01} are
\begin{equation}
\begin{cases}
\frac{d\hat{a}_{1}^{\dagger}}{dt} & =i\tilde{G}_{{\rm W}}\hat{m}-\frac{\kappa_{1}}{2}\hat{a}_{1}^{\dagger}+\sqrt{\kappa_{1}}(\hat{a}_{1}^{{\rm in}})^{\dagger},\\
\frac{d\hat{m}}{dt} & =-i\tilde{G}_{{\rm W}}^{*}\hat{a}_{1}^{\dagger}-\frac{\gamma}{2}\hat{m}+\sqrt{\gamma}\hat{m}^{{\rm in}},\\
\frac{d\hat{a}_{2}}{dt} & =-\frac{\kappa_{2}}{2}\hat{a}_{2}+\sqrt{\kappa_{2}}\hat{a}_{2}^{{\rm in}},
\end{cases}\label{eq:QLEWrite}
\end{equation}
where $\tilde{G}_{{\rm W}}=\alpha_{2}^{*}G_{21}^{+}$ is the cavity enhanced photon-magnon coupling, $\gamma$ is the magnon linewidth determined by the Gilbert damping coefficient of the material $\alpha_{{\rm Gilbert}}$ \citep{gilbertClassicsMagneticsPhenomenological2004}, and $\kappa_{1,2}$ are the cavity photon linewidths. We describe the open dynamics of the system through noise operators \citep{gardinerQuantumNoiseHandbook2000}. For the optical modes, $\hat{a}_{1,2}^{{\rm in}}$ describe vacuum fluctuations and the noise correlation function is given by
\begin{align}
\langle\hat{a}_{i}^{{\rm in}}(t)(\hat{a}_{i}^{{\rm in}})^{\dagger}(t^{\prime})\rangle & =\delta(t-t^{\prime}),\label{eq:NoisePhoton}\\
\langle(\hat{a}_{i}^{{\rm in}})^{\dagger}(t)\hat{a}_{i}^{{\rm in}}(t^\prime)\rangle & =0.
\end{align}
In turn, we assume that the magnon mode is coupled to a magnon thermal bath with mean number of quasiparticles $n_{{\rm Th}}$ described by the thermal noise operator $\hat{m}^{{\rm in}}$ satisfying
\begin{eqnarray}
\langle\hat{m}^{{\rm in}}(t)(\hat{m}^{{\rm in}})^{\dagger}(t^{\prime})\rangle & = & (n_{{\rm Th}}+1)\delta(t-t^{\prime}),\label{eq:NoiseMagnon}\\
\langle(\hat{m}^{{\rm in}})^{\dagger}(t)\hat{m}^{{\rm in}}(t^{\prime})\rangle & = & n_{{\rm Th}}\delta(t-t^{\prime}).
\end{eqnarray}
The mean number of magnons in the thermal bath $n_{{\rm Th}}$ and its temperature are related through the Bose-Einstein distribution $n_{{\rm Th}}=\left(\exp\left(\hbar\,\Omega/k_{B}T\right)-1\right)^{-1}$. 

Similarly for the read Hamiltonian \eqref{eq:BS01}, with $\tilde{G}_{{\rm R}}=\alpha_{1}^{*}G_{12}^{-}$,
\begin{equation}
\begin{cases}
\frac{d\hat{a}_{1}^{\dagger}}{dt} & =-\frac{\kappa_{1}}{2}\hat{a}_{1}^{\dagger}+\sqrt{\kappa_{1}}(\hat{a}_{1}^{{\rm in}})^{\dagger},\\
\frac{d\hat{m}}{dt} & =-i\tilde{G}_{R}\hat{a}_{2}-\frac{\gamma}{2}\hat{m}+\sqrt{\gamma}\hat{m}^{{\rm in}},\\
\frac{d\hat{a}_{2}}{dt} & =-i\tilde{G}_{R}^{*}\hat{m}-\frac{\kappa_{2}}{2}+\sqrt{\kappa_{2}}\hat{a}_{2}^{{\rm in}}.
\end{cases}\label{eq:QLERead}
\end{equation}
A schematic depiction of the linearization process is presented in Fig.~\ref{Linearization}. The solutions of Eqs. \eqref{eq:QLEWrite} and \eqref{eq:QLERead} have the form (for $X=$Write, Read)
\begin{equation}
\hat{\bm{A}}(t)=U^{X}(t)\cdot\hat{\bm{A}}(0)+{\displaystyle \int}_{0}^{t}d\tau U^{X}(t-\tau)\cdot\hat{\bm{N}}(\tau),\label{eq:SolutionsLangevin}
\end{equation}
where
\begin{eqnarray}
\hat{\bm{A}}=\left(\begin{array}{c}
\hat{a}_{1}^{\dagger}\\
\hat{m}\\
\hat{a}_{2}
\end{array}\right), &  & \hat{\bm{N}}=\left(\begin{array}{c}
\sqrt{k_{1}}(\hat{a}_{1}^{in})^{\dagger}\\
\sqrt{\gamma}\hat{m}^{in}\\
\sqrt{k_{2}}\hat{a}_{2}^{in}
\end{array}\right),\label{eq:DefinitionVectors}
\end{eqnarray}
and the evolution matrices $U^{{\rm Write}}(t)$ and $U^{{\rm Read}}(t)$ can be found analytically (see Appendix B). From now onwards we also assume $\kappa_{1}=\kappa_{2}=\kappa$.

For $\gamma\ll\kappa$ it is possible to retrieve the explicit state conditioned to the measurement of a write photon, by adiabatically eliminating the cavity dynamics together with considering suitably defined temporal modes (see Ref. \citep{gallandHeraldedSinglePhononPreparation2014}). For the material of choice in current experiments, YIG, the Gilbert damping parameter $\alpha_{{\rm Gilbert}}\approx10^{-4}$ \citep{tserkovnyakEnhancedGilbertDamping2002}, which is considered to be quite low compared to other magnetic materials. For magnon frequencies in the GHz range, $\gamma\sim$MHz while the WGM linewidth $\kappa$ can be of the order of GHz for the resonators used in current experiments \citep{zhangOptomagnonicWhisperingGallery2016}. Although the relative magnon linewidth $\gamma/\kappa$ in this case is small, thermalization processes are important for several applications, e.g., in the design of quantum memories \citep{afzeliusQuantumMemoryPhotons2015}. Moreover, for cavities of reduced size, in which the optomagnonic coupling could be enhanced \citep{violakusminskiyCoupledSpinlightDynamics2016}, detrimental effects on the quality factors for the optical fields due to both YIG patterning and confinement effects are expected \citep{grafCavityOptomagnonicsMagnetic2018a}. We therefore take into account both magnon and photon thermalization processes and consider the full treatment described above in terms of QLEs.

We set the initial conditions such that the state of the system at the beginning of the protocol $\rho(t=0)$ corresponds to the vacuum of both photon modes, plus a magnon thermal state with $n_{0}$ magnons:
\begin{eqnarray}
\rho(t=0) & = & \vert0\rangle\langle0\vert_{1}\otimes\vert0\rangle\langle0\vert_{2}\otimes\rho_{{\rm Th,m}},\nonumber \\
\rho_{{\rm Th,m}} & = & \frac{1}{1+n_{0}}\sum_{n\ge0}\left[\frac{n_{0}}{1+n_{0}}\right]^{n}\vert n\rangle\langle n\vert.\label{eq:InitState}
\end{eqnarray}
This state can be prepared as an equilibrium state with the environment, in which case $n_{0}=n_{{\rm Th}}$ [see Eq.~\eqref{eq:NoiseMagnon}], or by actively cooling, in which case $n_{0}$ is smaller than $n_{{\rm Th}}$. In our proposed setup, the latter can be realized by driving the read Hamiltonian Eq.~\eqref{eq:BS01} with a pump parameter $\alpha_{C}$ (related to the laser cooling power) from a state initially in equilibrium with the environment, until the system reaches a new thermalized steady state with $n_{0}<n_{{\rm Th}}$. The dissipative evolution will drive the system to a thermal state with mean number of magnons $n_{0}$ given by (see the Appendix B)
\begin{equation}
n_{0}=\frac{\gamma n_{{\rm Th}}}{(\kappa+\gamma)}\left(1+\frac{\kappa}{\gamma(1+4\alpha_{C}^{2}\,C)}\right),\label{eq:CoolingFormula}
\end{equation}
where $C=(G_{12}^{-})^{2}/\kappa\gamma$ is the single-photon cooperativity between magnons and $\hat{a}_{2}$ photons. The above cooling formula is valid under the linear Hamiltonian regime. For $\gamma \ll \kappa$ and $\alpha_{C}^{2}C\ll1$, this formula is equivalent to the one derived by Sharma \textit{et al.} in Ref. \citep{sharmaOpticalCoolingMagnons2018a}, not taking into account possible heating channels. For $4\alpha_{C}^{2}C\gg1$, the known result for strong coupled optomechanical systems is recovered \citep{liuDynamicDissipativeCooling2013}.

With the complete dynamics of the creation and annihilation operators given by Eq.~\eqref{eq:SolutionsLangevin} and the initial conditions set by Eq.~\eqref{eq:InitState}, we now proceed to characterize the heralding protocol in terms of expectation values of operators involving $\hat{a}_{1,2}^{(\dagger)}$ and $\hat{m}^{(\dagger)}$. The probability of measuring a photon during the write pulse can be retrieved via Mandel's formula \citep{gardinerQuantumNoiseHandbook2000}

\begin{widetext}

\begin{eqnarray}
P_{1,{\rm W}}(t) & = & \langle:\hat{a}_{1}^{\dagger}\hat{a}_{1}\exp(-\hat{a}_{1}^{\dagger}\hat{a}_{1}):\rangle \sim\langle\hat{a}_{1}^{\dagger}\hat{a}_{1}\rangle-\langle\hat{a}_{1}^{\dagger}\hat{a}_{1}^{\dagger}\hat{a}_{1}\hat{a}_{1}\rangle\label{Prob}\\
 & = & \frac{1}{8}-\frac{1}{8}\left[1-\frac{64\,\tilde{G}_{{\rm W}}^{2}}{F_{{\rm W}}^{2}}\left[(1+n_{0})\sinh^{2}\left(\frac{t}{4}F_{{\rm W}}\right)e^{-\frac{\kappa+\gamma}{2}t}+\frac{\gamma(1+n_{{\rm Th}})\mathcal{F}(t)}{(\kappa+\gamma)[(\kappa+\gamma)^{2}-F_{{\rm W}}^{2}]}\right]\right]^{2},\nonumber 
\end{eqnarray}
where $F_{{\rm W}}=\sqrt{(\kappa-\gamma)^{2}+16\tilde{G}_{{\rm W}}^{2}}$ and $\mathcal{F}(t)$ is given by
\begin{eqnarray*}
\mathcal{F}(t) & = & \left(1-e^{-\frac{\kappa+\gamma}{2}t}\right)F_{{\rm W}}^{2}-e^{-\frac{\kappa+\gamma}{2}t}(\kappa+\gamma)\left[F_{{\rm W}}\sinh\left(\frac{t}{2}F_{{\rm W}}\right)+2(\kappa+\gamma)\sinh^{2}\left(\frac{t}{4}F_{{\rm W}}\right)\right].
\end{eqnarray*}
\end{widetext}Equation \eqref{Prob} is a good approximation as long as $\langle\hat{a}_{1}^{\dagger}\hat{a}_{1}^{\dagger}\hat{a}_{1}^{\dagger}\hat{a}_{1}\hat{a}_{1}\hat{a}_{1}\rangle\ll\langle\hat{a}_{1}^{\dagger}\hat{a}_{1}^{\dagger}\hat{a}_{1}\hat{a}_{1}\rangle$. The temporal evolution is given in terms of two time scales $[\frac{(\kappa+\gamma)\pm F_{{\rm W}}}{2}]^{-1}$, which are controlled by the cooperativity $\tilde{C}_{{\rm W}}=\frac{\tilde{G}_{{\rm W}}^{2}}{\kappa\gamma}$. If $F_{{\rm W}}<(\kappa+\gamma)$, corresponding to $\tilde{C}_{{\rm W}}<1/4,$ the dynamics under the write pulse will drive the system to a steady state with a finite number of magnons and photons. Otherwise, if the cooperativity $\tilde{C}_{{\rm W}}>1/4$, the number of excitations in the system will grow exponentially in time. In the latter case, the dynamics will drive the system to a state with a high number of excitations in a short time interval. Hence, for the write phase a regime such that $\tilde{C}_{{\rm W}}<1/4$ should be aimed to, in order to keep $P_{1,{\rm W}}(t)$ and the mean number of photons and magnons small enough.

After the measurement (${\rm AM}$) of one write photon at $t_{{\rm m}}$, the expectation value of a given observable $\hat{X}(t)$ reads
\begin{equation}
\langle\hat{X}(t)\rangle_{{\rm AM}}=\frac{\langle\hat{a}_{1}^{\dagger}(t_{{\rm m}})\hat{X}(t)\hat{a}_{1}(t_{{\rm m}})\rangle}{\langle a_{1}^{\dagger}(t_{{\rm m}})a_{1}(t_{{\rm m}})\rangle},\label{eq:AfterMeasurement}
\end{equation}
which can be used to compute, for instance, the mean number of magnons in the heralded state $n_{{\rm hm}}=\langle\hat{m}^{\dagger}(t_{{\rm m}})\hat{m}(t_{{\rm m}})\rangle_{{\rm AM}}$. This is a heuristic model for the measurement used to study single photon sources and heralding \citep{razaviCharacterizingHeraldedSinglephoton2009}. A more accurate description of the photon measurement process can be given in terms of a stochastic model \citep{plenioQuantumjumpApproachDissipative1998a}, but this is beyond the purpose of this work. If a single magnon was successfully generated in the write phase, the read-photon field will be antibunched exhibiting a very low probability of a double photon measurement at a given instant. The (normalized) second-order correlation function of the read mode is given by
\begin{equation}
g_{{\rm Read}}^{(2)}(t,t+\tau)=\frac{\langle\hat{a}_{2}^{\dagger}(t)\hat{a}_{2}^{\dagger}(t+\tau)\hat{a}_{2}(t+\tau)\hat{a}_{2}(t)\rangle_{{\rm AM}}}{\langle\hat{a}_{2}^{\dagger}(t)\hat{a}_{2}(t)\rangle_{{\rm AM}}\langle\hat{a}_{2}^{\dagger}(t+\tau)\hat{a}_{2}(t+\tau)\rangle_{{\rm AM}}}.\label{eq:g2def}
\end{equation}
If the read photon is antibunched then the zero delayed correlation function fulfills $g_{{\rm Read}}^{(2)}(0)\equiv g_{{\rm Read}}^{(2)}(t,t)<1$, and $g^{(2)}(t,t+\tau)\ge g^{(2)}(t,t)$. On the other hand, for a thermal state $g_{{\rm Read}}^{(2)}(t,t)=2$ \citep{wallsQuantumOptics2008}. Experimentally, the second order correlation functions are measured via Hanbury-Brown-Twiss interferometry, a procedure also adopted for characterizing optomechanical heralding \citep{hongHanburyBrownTwiss2017,riedingerNonclassicalCorrelationsSingle2016}.

\section{Results}

We now proceed to quantify the results of the write and read protocol presented in the previous section. Following the parameters in current YIG-based optomagnonic systems, we set $\gamma/\kappa=10^{-2}$ corresponding to $\gamma\sim$ MHz and $\kappa\sim0.1$ GHz \citep{zhangOptomagnonicWhisperingGallery2016}. We also fix the time of the measurement at the end of the write pulse: $t_{{\rm m}}=T_{{\rm {\rm W}}}=10^{-7}$ s $\sim0.1/\gamma$ and $T_{\rm{off}}=T_{{\rm W}}/2=0.05/\gamma$, therefore satisfying the requirements imposed by Eq.~\eqref{eq:TimeConstraints}. The temporal width $T_{{\rm W}}$ adopted here for our calculations follows the one used in optomechanical heralding implementations (e.g., Ref. \citep{riedingerNonclassicalCorrelationsSingle2016}).

\subsection{Write Phase}

\begin{figure}
\begin{centering}
\includegraphics[width=1\columnwidth]{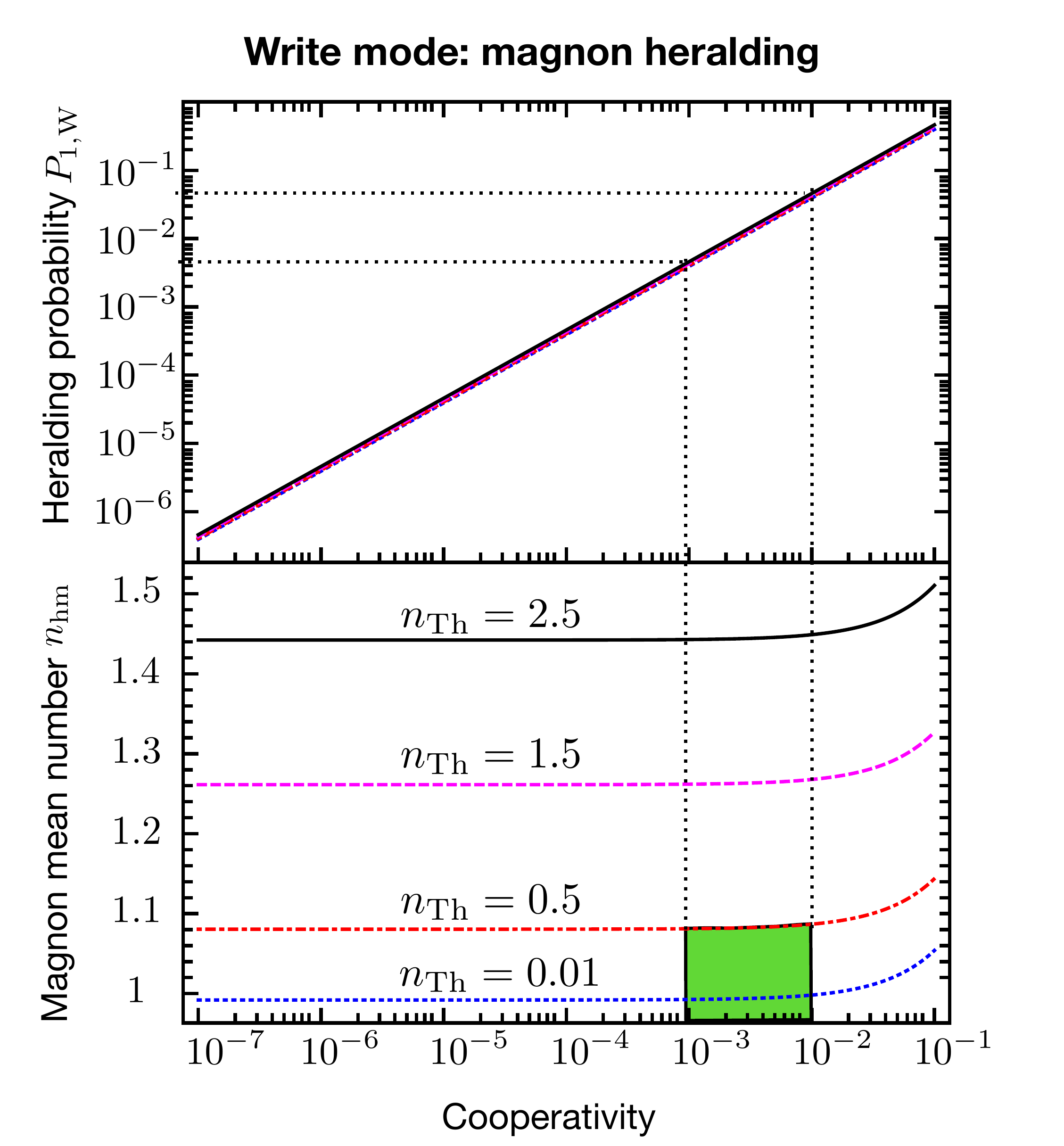}
\par\end{centering}
\caption{Probability $P_{1,{\rm W}}$ of measuring one write photon at time $t_{{\rm m}}$ (top) and mean number of magnons $n_{{\rm hm}}$ after the measurement of a write photon (bottom) as a function of the cooperativity $\tilde{C}_{{\rm W}}$ and mean number of thermal magnons $n_{{\rm Th}}$. $P_{1,{\rm W}}(t_{{\rm m}})$ is approximately linear with $\tilde{C}_{{\rm W}}$, while $n_{{\rm hm}}$ is only weakly sensitive to it for $\tilde{C}_{{\rm W}}\lesssim10^{-2}$. Although larger cooperativities enhance the probability of measuring a photon, there is a detrimental generation of more than one magnon indicating an imperfect Fock state. The shaded region indicates a combination of parameters for which a good one-magnon Fock state is generated while keeping the heralding probability appreciable. Temporal settings as given in the main text. }
\label{MeasProb}
\end{figure}

For the write phase two quantities are specially important: (i) the probability of measuring a write photon given in Eq.~\eqref{Prob}, which characterizes the probability of a heralding event, and (ii) the mean number of magnons $n_{{\rm hm}}$ after the measurement of a write photon. The value $n_{{\rm hm}}\sim1$ indicates the successful generation of a one-magnon Fock state. The probability of measuring a write photon $P_{1,W}$ at time $t_{{\rm m}}$ is depicted in Fig.~\ref{MeasProb} (upper plot) together with the corresponding mean number of magnons in the heralded state $n_{{\rm hm}}$ (bottom plot) as a function of the cooperativity of the write pulse $\tilde{C}_{{\rm W}}$. We considered different values of the magnon bath temperature, encoded in the parameter $n_{{\rm Th}}$, and we took the initial state of the system as the vacuum state ($n_{0}=0$). One sees that the probability of measuring one write photon grows linearly with the cooperativity $\tilde{C}_{{\rm W}}$ for the timescales considered, and it is weakly sensitive to the magnon temperature due to the small optomagnonic coupling. In contrast, the mean value of heralded magnons is almost independent of the cooperativity for $\tilde{C}_{{\rm W}}\lesssim10^{-2}$ (see bottom panel, Fig.~\ref{MeasProb}), whereas it has a relatively strong dependence on the bath temperature (see Appendix C). This behavior is due to an interplay between the optomagnonic coupling strength and the temporal width of the write pulse. For example, for shorter write pulses, $n_{{\rm hm}}$ is independent of $\tilde{C}_{{\rm W}}$ up to larger values of $\tilde{C}_{{\rm W}}$. As expected, the higher the temperature of the magnon bath, the higher the mean number of magnons in the heralded state. 

We can qualitatively understand the dependence of $P_{1,W}$ on $n_{\rm{Th}}$ by noticing that, since the optomagnonic coupling is small, the probability of a photon to be scattered by a magnon will depend only weakly on $n_{{\rm Th}}$. The same is true for understanding the dependence of $n_{{\rm hm}}$ on the cooperativity: since the optomagnonic coupling is small, the magnon dynamics for time scales comparable to $\gamma^{-1}$ will be influenced primarily by the interaction with the thermal bath.

Although an enhancement in the cooperativity $\tilde{C}_{{\rm W}}$ will improve the heralding probability, a detrimental effect is the generation of more than one magnon. There is therefore an interplay between the cooperativity and imperfections in the heralded state which needs to be tuned to achieve a given desired quality of the heralded state. For instance, in the green-marked region in Fig.~\ref{MeasProb}, the combination of parameters (cooperativity and temperature) is such that $n_{{\rm hm}}<$1.1, while the heralding probability is still appreciable, between $10^{-3}$ and $10^{-2}$. Additionally, even for small number of thermal magnons, imperfections in the Fock state are expected according to the duration of the write phase. Thermal effects can be minimized by shorter write pulses, leading consequently to shorter measurement times $t_{{\rm m}}$; however, the heralding probability also decreases.

In the strong-coupling regime, the write pulse will create a huge number of excitations unless its duration is $T_{{\rm W}}\ll\vert\tilde{G}_{{\rm W}}\vert$. On the other hand, for the read phase, strong coupling allows coherent mapping between magnons and photons. In the following we study the read phase in both weak- and strong-coupling limits, while fixing the cooperativity of the write phase at $\tilde{C}_{{\rm W}}=10^{-2}$, in correspondence with the discussion in the last paragraphs.

\subsection{Read phase for weak coupling}

\begin{figure}
\centering{}\includegraphics[width=1\columnwidth]{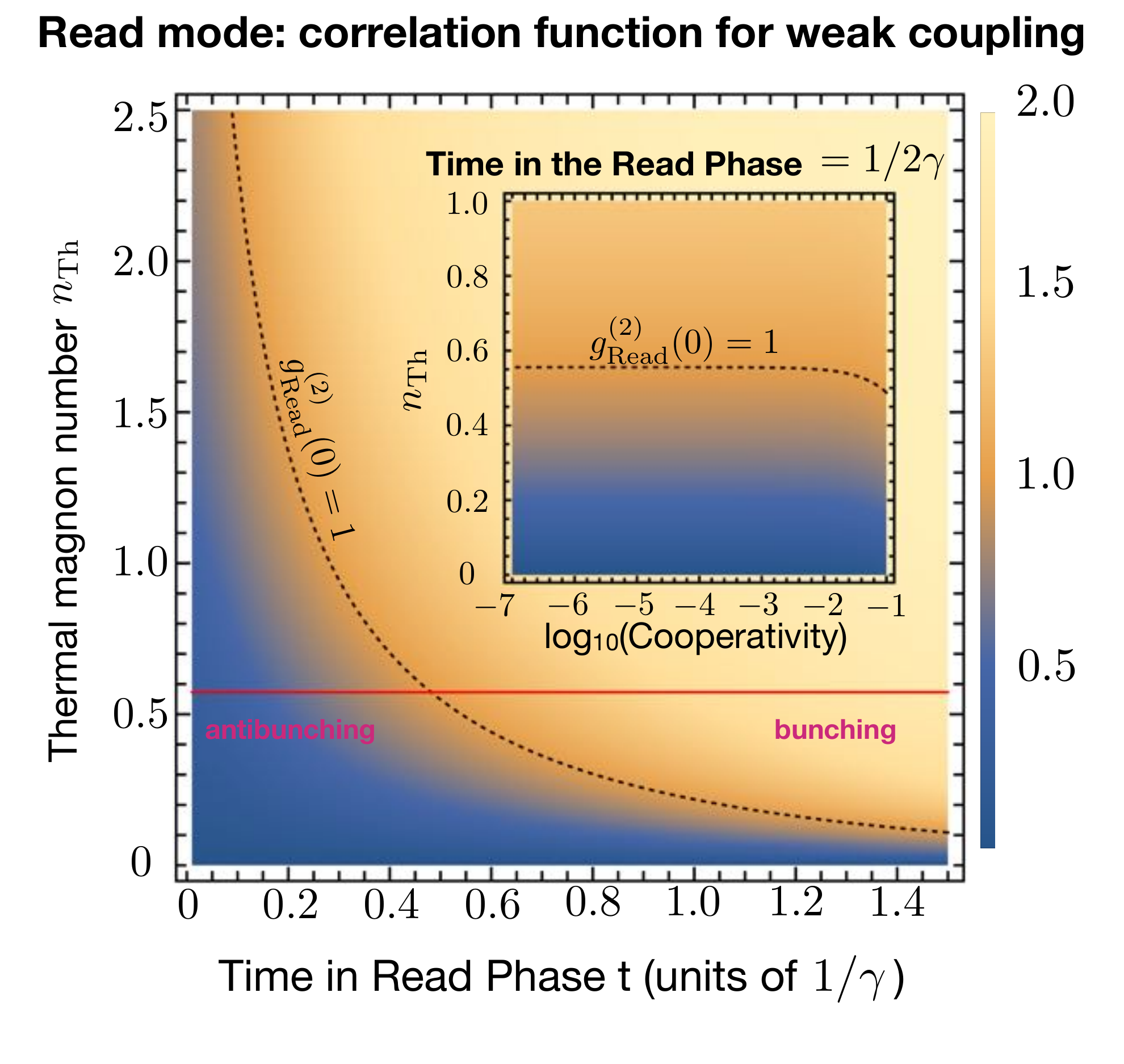}
\caption{Second-order correlation function at zero delay $g_{{\rm Read}}^{(2)}(0)$ of the read photon conditioned to the measurement of a write photon. $g_{{\rm Read}}^{(2)}(0)$ is shown as a function of the mean number of thermal magnons $n_{{\rm Th}}$ and of the time in the read phase (in units of $1/\gamma$ and $t=0$ corresponds to the beginning of the read phase) for $\tilde{C}_{{\rm W}}=\tilde{C}_{R}=10^{-2}$. The black dashed line marks the transition between antibunching and bunching behaviors. The red line indicates the threshold for which the number of heralded magnons $n_{{\rm hm}}<1.1$. Below this line antibunching of the correlation function indicates a successful Fock state generation. Inset: $g_{{\rm Read}}^{(2)}(0)$ as a function of $\rm{log}_{10}(\tilde{C}_{{\rm R}})$. Temporal settings as in the main text. }
\label{SecOrdFirst}
\end{figure}

The weak-coupling regime $\vert\tilde{G}_{{\rm R}}\vert<\kappa$ implies $\tilde{C}_{{\rm R}}=\vert\tilde{G}_{{\rm R}}\vert^{2}/\kappa\gamma<\kappa/\gamma\sim10^{2}$. To characterize the read phase we use the zero-delay correlation function $g_{{\rm Read}}^{(2)}(0)$ given by Eq.~\eqref{eq:g2def}. Figure \ref{SecOrdFirst} shows the results for $g_{{\rm Read}}^{(2)}(0)$ for a fixed cooperativity $\tilde{C}_{{\rm W}}=\tilde{C}_{{\rm R}}=10^{-2}$ as a function of time during the read phase, and as a function of the mean number of thermal magnons $n_{{\rm Th}}$. We have indicated the line corresponding to $g_{{\rm Read}}^{(2)}(0)=1$ as a visual guide, marking the transition between antibunching ($g_{{\rm Read}}^{(2)}(0)<1$) to bunching ($g_{{\rm Read}}^{(2)}(0)>1$). This transition depends on the magnon bath temperature, exhibiting two distinct behaviors. For small temperatures, the evolution from bunching to antibunching depends on the magnon thermalization characteristic time $1/(n_{{\rm Th}}\gamma)$, which can be long for temperatures approaching zero. For larger temperatures the transition to $g_{{\rm Read}}^{(2)}(0)>1$ is faster and dominated by the photon's decay rate. In this last case, the magnon state is closer to a thermal state rather than to a Fock state, and the bunching to antibunching dynamics reflects the interaction of the photon field with such a thermal magnon state. Therefore, antibunching for such large magnon bath temperatures is not a signal of successful heralding; see the red horizontal line in Fig.~\ref{SecOrdFirst}. The inset shows the weak dependence of $g_{{\rm Read}}^{(2)}(0)$ on the cooperativity of the read pulse $\tilde{C}_{{\rm R}}$. We notice that a non trivial dependence on the cooperativity appears for $\tilde{C}_{{\rm R}}>10^{-2}$ as a consequence of the increased coupling that compensates the decay processes, similar to the discussion in the last section referring to the write phase. 

\subsection{Read phase for strong coupling}

The reading protocol is limited by the strength of the coupling between mode 2 and the heralded magnon. The current experimental state of the art in cavity optomagnonics is that of systems in the weak-coupling regime \citep{zhangOptomagnonicWhisperingGallery2016,osadaOrbitalAngularMomentum2018,haighTripleResonantBrillouinLight2016,osadaCavityOptomagnonicsSpinOrbit2016}. This is, however, a very young field, and it can be expected that the current values of the coupling and cooperativities will be improved in next generation experiments. The strong-coupling regime is highly appealing, since it is a prerequisite for many quantum protocols. In our case, the strong-coupling regime for the read phase leads to Rabi oscillations between the magnon and photon fields, allowing for coherent state transfer \citep{verhagenQuantumcoherentCouplingMechanical2012,palomakiCoherentStateTransfer2013}.

\begin{figure}
\centering{}\includegraphics[width=1\columnwidth]{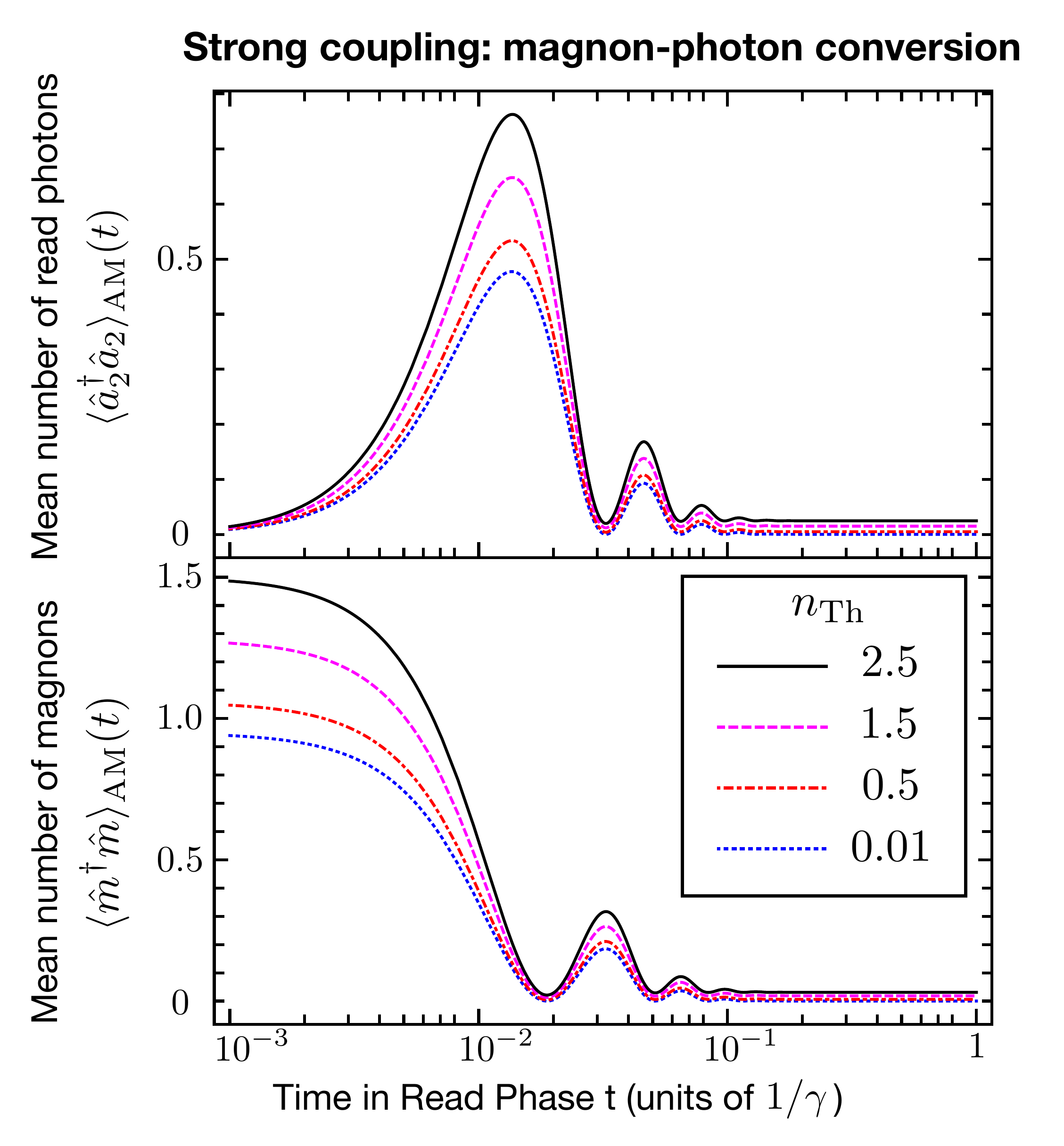}
\caption{Oscillations between read photons and magnons in the strong-coupling regime. The plots depict the mean number of read photons (top) and the mean number of magnons (bottom) as a function of the time during the read phase, for a read cooperativity $\tilde{C}_{{\rm R}}=100$, corresponding to $\tilde{G}_{{\rm R}}\sim\kappa$. The different curves correspond to different  temperatures of the magnon bath.}
\label{FigUltraStrong01}
\end{figure}

For high cooperativities, fast oscillations between read photons and magnons take place. For the parameters adopted in the last section, the damped oscillations between the mean number of read photons and the mean number of magnons have oscillation periods of $\sim10^{-8}$s, as depicted in Fig.~\ref{FigUltraStrong01} for strong optomagnonic coupling $\tilde{G}_{{\rm R}}=\kappa\sim0.1$ GHz, corresponding to a cooperativity of $\tilde{C}_{{\rm R}}\sim10^{2}$. Although this cooperativity value is quite large with respect to the state of the art in YIG-based optomagnonics ( $\sim10^{-7}$), in cold atoms cavity systems the strong-coupling regime is attainable \citep{kohlerCavityAssistedMeasurementCoherent2017,welteCavityCarvingAtomic2017,kohlerNegativeMassInstabilitySpin2018}.

\begin{figure}
\centering{}\includegraphics[width=1\columnwidth]{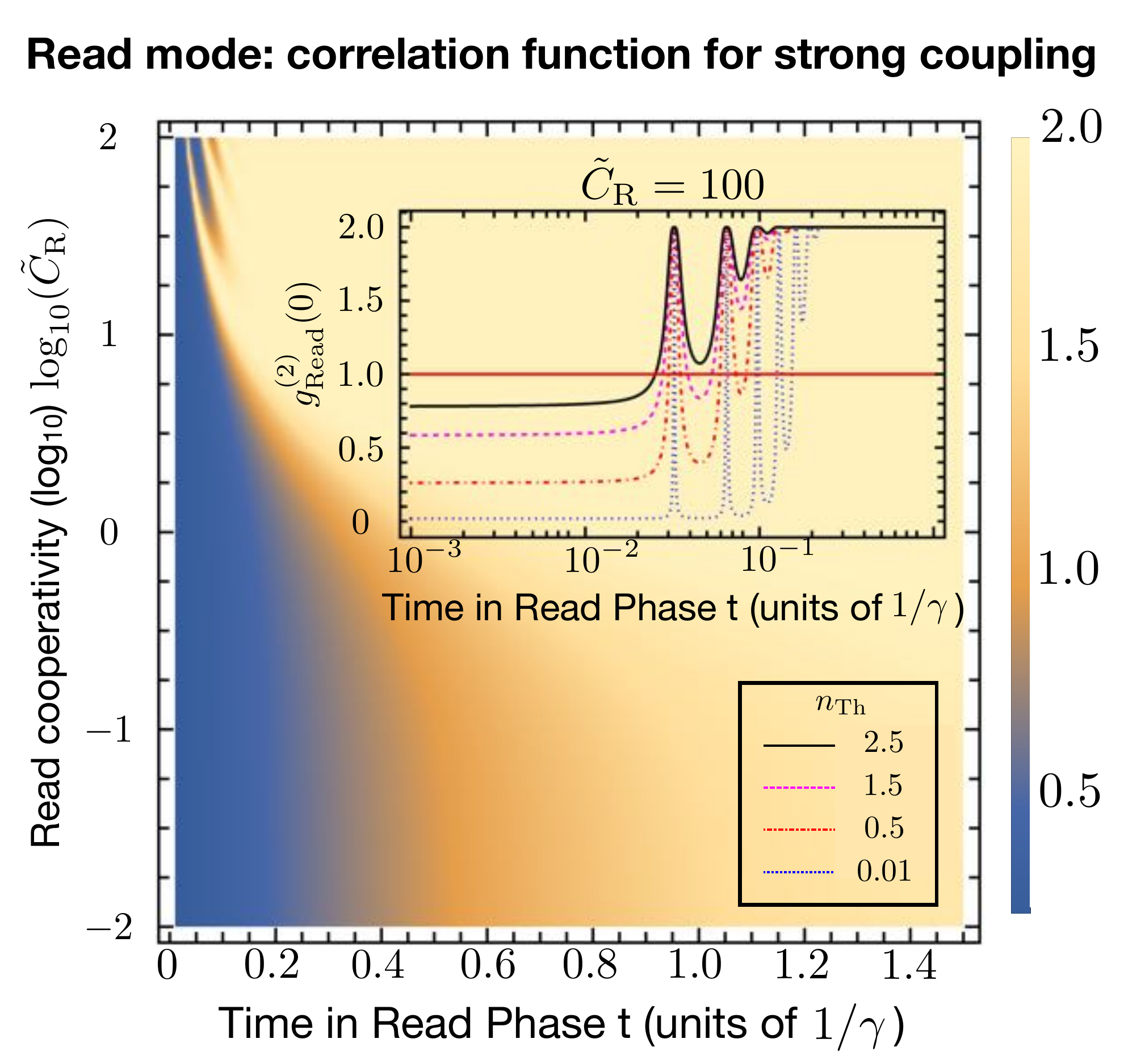}
\caption{Second-order correlation function $g_{{\rm Read}}^{(2)}(0)$ of the read photon mode in the strong-coupling regime. $g_{{\rm Read}}^{(2)}(0)$ is plotted as a function of the time in the read phase and of the logarithm of the read phase cooperativity $\log_{10}(\tilde{C}_{\rm{R}})$ [related to the pumping parameter $\alpha_{1}$ see Eq.~\eqref{eq:BS01}] for $n_{{\rm Th}}=0.5$. The inset depicts the temporal evolution of the correlation function for $\tilde{C}_{R}=100$ and for different temperatures of the magnon bath. The oscillations indicate coherent Rabi oscillations between photons and magnons.}
\label{FigureStrong02}
\end{figure}

As shown in Fig.~\ref{FigureStrong02}, the fast oscillations between magnons and read photons are also visible in the second order correlation function of the read mode. Analogously to the mean number of photons, the correlation function rapidly oscillates in time, encoding the coherent magnon-photon oscillations. These temporal oscillations in the correlation function could be resolved either with state of the art Hanbury-Brown and Twiss interferometers (see, e.g, Refs. \citep{hongHanburyBrownTwiss2017,riedingerRemoteQuantumEntanglement2018}) or other devices; see, e.g., Ref. \citep{g2measurement}, where an interferometric setup is implemented in a semiconductor-superconductor platform.

We emphasize that the interpretation of $g^{(2)} (0)$ as a witness of successfully heralding a magnon Fock state depends on the temperature of the magnon mode. From the experimental point of view this requires knowledge of the mean number of thermal magnons $n_{\rm{Th}}$. Other possible witnesses could be cross-correlation functions involving both the magnon and the photon mode, such as $$g^2_{\rm{m}, 2} = \frac{\langle\hat{a}_2^\dagger \hat{m}^\dagger \hat{a}_2 \hat{m} \rangle_{\rm{AM}}}{\langle \hat{m}^\dagger \hat{m} \rangle_{\rm{AM}} \langle \hat{a}^\dagger_2 \hat{a}_2 \rangle_{\rm{AM}}}.$$These also have bounds which are violated by nonclassical states and can be measured with interferometric techniques involving only the optical modes. This type of cross-correlation function has been also used to witness the generation of phonon Fock states \citep{riedingerNonclassicalCorrelationsSingle2016}, although they suffer from the same limitations as $g^{(2)}(0)$. In a more complex setup, in which the magnons are also strongly coupled to a qubit in the dispersive regime, it is possible to resolve the quanta of the magnon mode via a spectroscopic measurement of the qubit \citep{lachance-quirionResolvingQuantaCollective2017b}. Although this setup also requires small temperatures, it does not require the precise knowledge of $n_{\rm{Th}}$.

\subsection{Effects of the initial state and cooling requirements}

\begin{figure}
\centering{}\includegraphics[width=1\columnwidth]{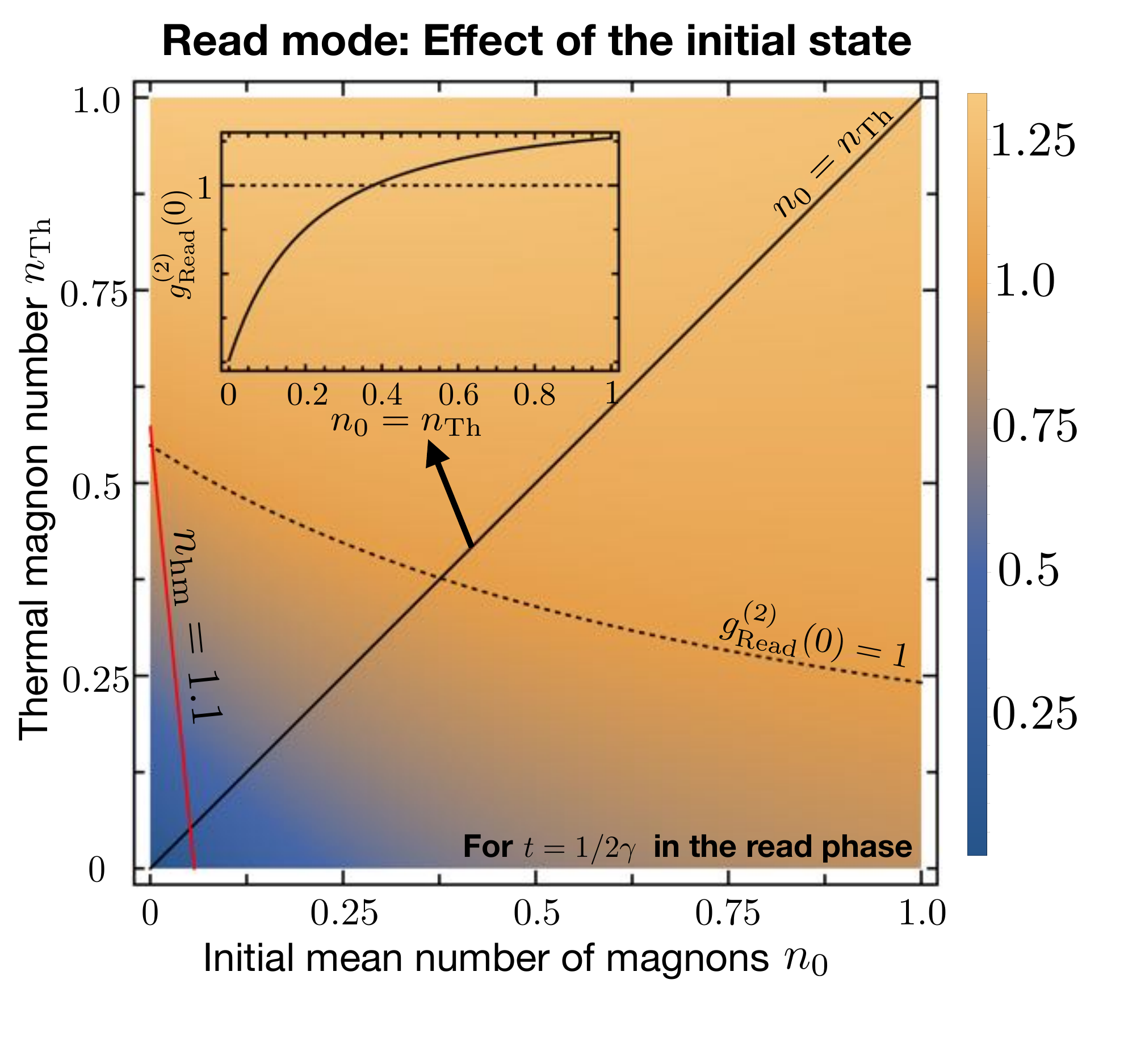}
\caption{Effects of the initial state on the autocorrelation function $g_{{\rm Read}}^{(2)}(0)$ of the read photon, for an active-cooling setup (region above the line $n_{0}=n_{{\rm Th}}$ ) and for an initial state in thermal equilibrium with the bath (inset). The red line indicates the threshold for which $n_{{\rm hm}}<1.1$. Results for $\tilde{C}_{{\rm W}}=\tilde{C}_{{\rm R}}=10^{-2}$, $t_{{\rm m}}=0.1/\gamma$, and $t=1/2\gamma$ in the read phase.}
\label{FigureInitState}
\end{figure}

The initial state has a strong effect on the heralding protocol. Figure \ref{FigureInitState} depicts $g_{{\rm Read}}^{(2)}(0)$ at $t=1/2\gamma$ after the beginning of the read phase, as a function of the initial state mean number of magnons $n_{0}$ and of the thermal magnon number $n_{{\rm Th}}$. An active-cooling setup corresponds to the region above the black continuous line for which $n_{0}<n_{{\rm Th}}$ (note that $n_{0}$ is the initial state for the heralding protocol, after cooling). The inset shows $g_{{\rm Read}}^{(2)}(0)$ at $t=1/2\gamma$ for an initial thermal state $n_{0}=n_{{\rm th}}$, and the red line indicates the region for which the number of heralded magnons is $n_{{\rm hm}}<1.1$. We notice that even when the mean number of magnons in the state after the measurement is not close to one, it is possible that $g_{{\rm Read}}^{(2)}(0)<1$. This is again due to the interaction of the photon mode with a imperfect Fock magnon state, which limits the bunching to antibunching transition by the photon lifetime as in the plot of Fig.~\ref{SecOrdFirst}, and makes the autocorrelation function of the photon field an unreliable witness for the successfulness of the magnon heralding. 

Since the initial state imposes tight constraints on the magnon heralding, in particular having a strong influence on the mean number of magnons in the system after the measurement of a write photon, we study the cooling requirements to obtain a mean number of heralded magnons $n_{{\rm hm}}\le1+\epsilon$, where $\epsilon$ is a tolerance margin with respect to the ideal case $n_{{\rm hm}}=1$. This is indirectly related to the fidelity of the heralded state with respect to the perfect one-magnon Fock state (see Appendix E) and sets a minimum value for the power of the cooling laser. 

\begin{figure}
\centering{}\includegraphics[width=1\columnwidth]{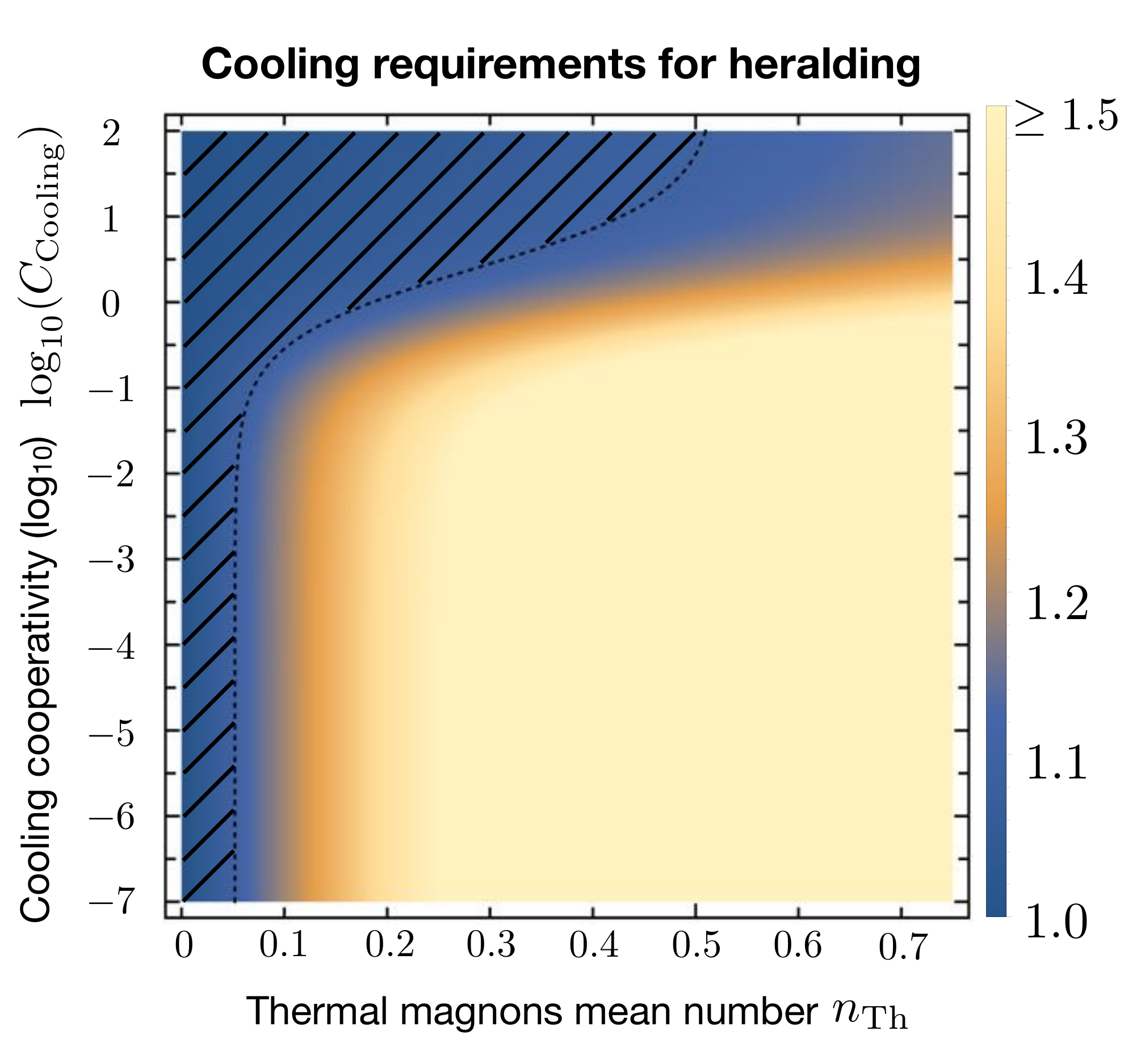}
\caption{Mean number of heralded magnons $n_{{\rm hm}}$ (for $t_{{\rm m}}=0.1\gamma^{-1}\sim10^{-7}$ s) as a function of the logarithm of the cooling cooperativity $C_{{\rm Cooling}}=\alpha_{C}^{2}\,C$ and of the average number of magnons in the thermal bath. The shaded area indicates the region for which $n_{{\rm hm}}\le1.1$, corresponding to a tolerance margin of $0.1$ with respect to the ideal case.}
\label{CoolingPower}
\end{figure}

An initial state in the form of Eq.~\eqref{eq:InitState} with mean number of magnons $n_{0}$ is obtained by actively cooling a thermal state with mean number of magnons $n_{{\rm Th}}$. For such a state, $n_{0}$ (such that $n_{0}<n_{{\rm Th}}$) is given by the cooling formula Eq.~\eqref{eq:CoolingFormula} and determined by $n_{{\rm Th}}$, $\gamma/\kappa$, and by the cooling cooperativity $C_{{\rm Cooling}}=\alpha_{C}^{2}C$. The tolerance margin $\epsilon$ imposed on $n_{{\rm hm}}$ translates into an upper limit for $n_{0}$, therefore imposing constraints on $C_{{\rm Cooling}}$. Figure \ref{CoolingPower} shows $n_{{\rm hm}}$ as a function of $n_{{\rm Th}}$ and of $\log_{10}(C_{{\rm Cooling}})$ for an initial state obtained by active cooling. The shaded area indicates the region for which $n_{{\rm hm}}\le1.1$, corresponding to a tolerance margin of $\epsilon=0.1$. We see that the tolerance margin may not be attainable depending on the value of $n_{{\rm Th}}$ for a given $\gamma/\kappa$. The smaller the ratio $\gamma/\kappa$, the more $n_{0}$ can be made closer to zero, so that the same tolerance margin can be attained with smaller cooperativities and for higher temperatures.

For the tolerance margin $\epsilon=0.1$, taking $\gamma/\kappa=10^{-2}$ and a magnon mode frequency $\Omega\sim$ 10 GHz, the maximum possible temperature $T_{\max}$ in order to perform the protocol without active cooling (i.e., for an initial state in thermal equilibrium with the magnon bath, $n_{0}=n_{{\rm Th}}$) is $T_{\max}\sim25.38\,{\rm mK}$, corresponding to $n_{{\rm Th}}=0.025$. In that case the fidelity of the heralded state to a one-magnon Fock state is $0.943$. Otherwise, for the same parameters but considering a bath at temperature  $T_{{\rm bath}}=\{30,20,50\}$ mK ($n_{{\rm Th}}=\{0.085,0.174,0.277\}$), the required cooling cooperativity for the same tolerance margin is $C_{{\rm Cooling}}\gtrsim\{0.191,0.882,2.317\}$, respectively. The corresponding fidelities of the heralded state with respect to the one-magnon Fock state are $\{0.943,942,0.941\}$, which are higher than the estimate fidelity of heralded phonon states in optomechanics experiments $\sim0.88$ \citep{riedingerNonclassicalCorrelationsSingle2016}. As already pointed out in Sec. V, these cooperativity values are quite large compared to current values in YIG-based cavity optomagnonic systems. Improvement in sample design can enhance the magnon-photon cooperativities as to achieve the required values for the protocol \citep{grafCavityOptomagnonicsMagnetic2018a,zhangOptomagnonicWhisperingGallery2016}. Moreover, samples with better magnon linewidth would require smaller cooperativites to achieve the same tolerance margin. For the examples above but now considering $\gamma/\kappa=10^{-4}$, the necessary cooperativities are $\gtrsim\{0.175,0.622,1.142\}$. In the limit $\gamma/\kappa=0$ any thermal state can be cooled to its ground state. 

\section{Conclusions}

We proposed a magnon heralding protocol in a cavity optomagnonics setup, in which the measurement of one optical photon collapses an entangled magnon-photon state to a single-magnon Fock state. Our starting point was a linearized optomagnonic Hamiltonian with an asymmetric coupling between one magnon mode and two non-degenerate photon modes. This model is in correspondence with recent experiments with YIG spheres and takes into account specific selection rules involving conservation of energy and angular momentum.

The linearized optomagnonic Hamiltonian includes resonances in the form of a two-mode parametric amplifier (``write'') and beam-splitter (``read'') type interactions between the photon modes and the magnon mode. Considering an initial state close to the vacuum of the system, the protocol is implemented by first driving the write interaction, which generates pairs of correlated photon-magnon states. The subsequent measurement of a ``write'' photon collapses the state of the system to a single-magnon Fock state, which can then be mapped to a ``read'' photon mode by driving the beam-splitter interaction. The nonclassicality of the state can be certified by measuring the autocorrelation of the photon field.

An important figure of merit in these systems is the cooperativity, which measures the strength of the coupling relative to the dissipation channels for magnons and photons. We showed that the heralding quantum protocol for magnon Fock states can be realized with cooperativities of the order of $10^{-2}$ if the system is cooled to its ground state. Although this requires an improvement with respect to the state of the art cooperativity in solid state optomagnonic systems, it is to be expected that new generation experiments will reach this threshold \citep{zhangOptomagnonicWhisperingGallery2016,grafCavityOptomagnonicsMagnetic2018a}. Provided cooperativities of this order can be achieved, and considering magnonic and optical linewidths consistent with current experiments (see appendix D for a short analysis for different linewidth regimes) the obtained heralding probability is in line with the ones realized in recent optomechanical experiments \citep{mcconnellEntanglementNegativeWigner2015,riedingerNonclassicalCorrelationsSingle2016,riedingerRemoteQuantumEntanglement2018}. Achieving the strong-coupling regime would moreover allow the conversion of the heralded state to a photon mode by the read part of the protocol. 

We showed that, whereas small deviations of the initial state from the magnonic vacuum have a strong detrimental impact on the heralding protocol, this can be circumvented via actively cooling the system prior to the write phase. Accordingly, we derived cooling requirements to generate one-magnon Fock states with high fidelity. 

In the main text of this manuscript we presented an analytical analysis based in the solutions of quantum Langevin equations for square light pulses. We have also performed a numerical analysis taking into account more realistic pulse shapes (in the form of Gaussian pulses), which are presented in the appendix F. These numerical results point to the same overall conclusions, with some additional detrimental effects due to the larger timescales involved. 

The heralding protocol proposed here can be the basis for implementing hybrid quantum information processing schemes \citep{andersenHybridDiscreteContinuousvariable2015} with cavity optomagnonics. Moreover, such heralding protocol could be used as the starting point to explore quantum mechanics with macroscopic systems, since magnon Fock states are truly nonclassical states involving a huge collection of solid state spins. Finally, the selection rules of the optomagnonic coupling generates polarization-dependent processes \citep{osadaCavityOptomagnonicsSpinOrbit2016} that could be interfaced with chiral circuits \citep{lodahlChiralQuantumOptics2017} to engineer devices such as quantum isolators and circulators. The heralding protocol in such a setup would be a preparation step and the overall function of the device would be similar to those implemented in atomic systems \citep{petersenChiralNanophotonicWaveguide2014,shomroniAllopticalRoutingSingle2014}. 

\begin{acknowledgments}
The authors thank Carlos Navarrete-Benlloch, Vittorio Peano, and Florian Marquardt for useful discussions and acknowledge financial support from the Max Planck Gesellschaft through an Independent Max Planck Research Group. 
\end{acknowledgments}

\section*{Appendix A - Linearization of the optomagnonic Hamiltonian}

In order to describe the dynamics of the magnon-photon system in the heralding protocol, we consider the linearized version of the pumped optomagnonic Hamiltonian obtained by adding to Eq.~\eqref{eq:OMHamiltonian-1} a driving term 
\begin{equation}
\hat{H}_{{\rm tot}}=\hat{H}+\hat{H}_{{\rm driving}},\label{eq:TotalDriving-1}
\end{equation}
\[
\hat{H}_{{\rm driving}}=i\epsilon_{i}(\hat{a}_{i}e^{i\omega_{L}t}-\hat{a}_{i}^{\dagger}e^{-i\omega_{L}t}),
\]
where $i$ = 1 or 2 indicates the pumped mode, $\omega_{L}$ is the laser frequency and $\epsilon_{i}=\hbar\sqrt{\frac{2\kappa_{i}\mathcal{P}_{i}}{\hbar\omega_{L}}}$ depends on the driving laser power $\mathcal{P}_{i}$ and on the coupling between the mode and the fiber $\kappa_{i}$.

Considering from now on the framework of the paper, involving two photon modes and one magnon mode, first the time dependence of the pumping term is eliminated through $\hat{H}\rightarrow\hat{U}\hat{H}\hat{U}^{\dagger}-i\hbar\hat{U}\frac{\partial\hat{U}^{\dagger}}{\partial t}$, where $\hat{U}=\exp(-i\omega_{L}t(\hat{a}_{1}^{\dagger}\hat{a}_{1}+\hat{a}_{2}^{\dagger}\hat{a}_{2}))$ as to have
\begin{eqnarray}
\hat{H} & = & -\hbar\Delta_{1}\hat{a}_{1}^{\dagger}\hat{a}_{1}-\hbar\Delta_{2}\hat{a}_{2}^{\dagger}\hat{a}_{2}\nonumber \\
 & + & {\displaystyle \hbar\Omega\hat{m}^{\dagger}\hat{m}+\hbar}\left(G_{12}^{-}\hat{a}_{1}^{\dagger}\hat{a}_{2}+G_{21}^{-}\hat{a}_{1} \hat{a}_{2}^{\dagger} \right)\hat{m}^\dagger+h.c.\nonumber \\
 & + & i\epsilon_{i}(\hat{a}_{i}^{\dagger}-\hat{a}_{i}).
\end{eqnarray}

The Langevin equations for the operators of the photon and the magnon modes are thus

\begin{eqnarray*}
\frac{d\hat{a}_{1}}{dt} & = & i\Delta_{1}\hat{a}_{1}-iG_{12}^{-}{\displaystyle \hat{a}_{2}}\hat{m}^{\dagger}-i(G_{21}^{-})^* \hat{a}_{2}\hat{m}+\epsilon_{i} \delta_{i,1}-\frac{\kappa_{1}}{2}\hat{a}_{1},\\
\frac{d\hat{a}_{2}}{dt} & = & i\Delta_{2}\hat{a}_{2}-i(G_{12}^{-})^{*}{\displaystyle \hat{a}_{1}}\hat{m}-iG_{21}^{-}\hat{a}_{1}\hat{m}^{\dagger}+\epsilon_{i} \delta_{i,2}-\frac{\kappa_{2}}{2}\hat{a}_{2},\\
\frac{d\hat{m}}{dt} & = & -i\Omega\hat{m}-iG_{12}^{-}\hat{a}_{1}^{\dagger}\hat{a}_{2}-iG_{21}^{-}\hat{a}_{1}\hat{a}_{2}^{\dagger}-\frac{\gamma}{2}\hat{m}.
\end{eqnarray*}
The steady state values $\langle\,\hat{a}_{i}\,\rangle=\alpha_{i}$ and $\langle\hat{m}\rangle=\beta$ are thus given by the set of nonlinear equations

\begin{eqnarray*}
\left(i\tilde{\Delta}_{1}-\frac{\tilde{\kappa}_{1}}{2}\right)\alpha_{1} & = & -\epsilon_{i} \delta_{i,1},\\
\left(i\tilde{\Delta}_{2}-\frac{\tilde{\kappa}_{2}}{2}\right)\alpha_{2} & = & -\epsilon_{i} \delta_{i,2},\\
\beta & = & -\frac{i\left(\alpha_{1}^{*}\alpha_{2}G_{12}^{-}-G_{21}^{-}\alpha_{1}\alpha_{2}^{*}\right)}{i\Omega+\gamma/2},
\end{eqnarray*}
where, considering $\Omega\gg\kappa_{1,2},\gamma$, 
\begin{eqnarray*}
\tilde{\Delta}_{i} & = & \Delta_{i}+\frac{\Omega\left(\vert G_{12}^{-}\vert^{2}-\vert G_{21}^{-}\vert^{2}\right)\vert\alpha_{j}\vert^{2}}{\Omega^{2}+\gamma^{2}/4},(i\neq j)\\
\tilde{\kappa}_{1} & = & \kappa_{1}+\frac{\gamma\left(\vert G_{12}^{-}\vert^{2}+\vert G_{21}^{-}\vert^{2}\right)\vert\alpha_{2}\vert^{2}}{\Omega^{2}+\gamma^{2}/4},\\
\tilde{\kappa}_{2} & = & \kappa_{2}-\frac{\gamma\left(\vert G_{12}^{-}\vert^{2}+\vert G_{21}^{-}\vert^{2}\right)\vert\alpha_{1}\vert^{2}}{\Omega^{2}+\gamma^{2}/4}.
\end{eqnarray*}

Since only the mode $i=1$ or $2$ is pumped, then $\alpha_{j}=0$ if $j\neq i$, $\beta=0$, and $\tilde{\Delta}_{P}=\Delta_{P}$ and for the pumped mode 
\[
\alpha_{i}=-\frac{\epsilon_{i}}{i\Delta_{i}-\kappa_{i}/2}.
\]
The linearized Hamiltonian is obtained by considering fluctuations around such coherent state solutions through the displacement $\hat{a}_{i}\rightarrow\alpha_{i}+\hat{a}_{i}$, $\hat{m}\rightarrow\hat{m}$ (since its steady-state mean value is $\beta=0$). Discarding nonlinear terms one obtains
\begin{eqnarray}
\hat{H} & = & -\sum_{i=1,2}\hbar\Delta_{i}\hat{a}_{i}^{\dagger}\hat{a}_{i}+\hbar\Omega\hat{m}^{\dagger}\hat{m}\nonumber \\
 &  & +\hbar\alpha_{1}^{*}\hat{a}_{2}\left(G_{12}^{+}\hat{m}+G_{12}^{-}\hat{m}^{\dagger}\right)+h.c.\nonumber \\
 &  & +\hbar\alpha_{2}\hat{a}_{1}^{\dagger}\left(G_{12}^{+}\hat{m}+G_{12}^{-}\hat{m}^{\dagger}\right)+h.c.
\end{eqnarray}
The interaction-picture version of this Hamiltonian is Eq.~\eqref{eq:TimeDependentHamiltonian-1-1}, which we considered in our calculations.

\section*{Appendix B - Temporal evolution matrices and cooling formula}

The linear quantum Langevin equations describing the temporal evolution of the field operators during the write and read phases are straightforwardly solved. By writing the Langevin equations for a given phase $P=W,$ $R$ as [see Eq.~\eqref{eq:DefinitionVectors}]
\[
\frac{d\hat{\bm{A}}}{dt}=M^{P}\cdot\hat{\bm{A}}(t)+\bm{\hat{N}}(t),
\]
we perform a basis transformation $X_{P}$ such that $\left(M^{{\rm P}}\right)^{\prime}=(X_{{\rm P}})^{-1}M^{{\rm P}}X_{{\rm P}}={\rm diag}\{\lambda_{1}^{{\rm P}},\lambda_{2}^{{\rm P}},\lambda_{3}^{{\rm P}}\}$. Thus the components of the Langevin equations in such basis read
\[
\frac{d\hat{A}_{i}^{\prime}}{dt}=\lambda_{i}^{{\rm P}}\hat{A}_{i}^{\prime}(t)+\hat{N}_{i}^{\prime}(t),
\]
which can be integrated as to have 
\[
\hat{A}_{i}^{\prime}(t)=e^{\lambda_{i}^{{\rm P}}t}\hat{A}_{i}^{\prime}(0)+\int_{0}^{t}d\tau\,e^{\lambda_{i}^{{\rm P}}(t-\tau)}\hat{N}_{i}^{\prime}(\tau).
\]
The solutions depicted in Eq.~\eqref{eq:SolutionsLangevin} are obtained by transforming back to the original basis $\hat{\bm{A}}(t)=X_{{\rm P}}\hat{\bm{A}}^{\prime}(t)$. The the time evolution matrices $U^{{\rm Write}}(t)$ and $U^{{\rm Read}}(t)$ are given by\begin{widetext}

\begin{eqnarray*}
U^{{\rm Write}}(t) & = & \left(\begin{array}{ccc}
\frac{e^{-(\kappa+\gamma)t/4}}{F_{{\rm W}}}C_{-}^{{\rm W}} & \frac{4i\tilde{G}_{{\rm W}}}{F_{{\rm W}}}e^{-(\kappa+\gamma)t/4}\sinh\left[\frac{tF_{{\rm W}}}{4}\right] & 0\\
-\frac{4i\tilde{G}_{{\rm W}}}{F_{{\rm W}}}e^{-(\kappa+\gamma)t/4}\sinh\left[\frac{tF_{{\rm W}}}{4}\right] & \frac{e^{-(\kappa+\gamma)t/4}}{F_{{\rm W}}}C_{+}^{{\rm W}} & 0\\
0 & 0 & e^{-\kappa t/2}
\end{array}\right),\\
U^{{\rm Read}}(t) & = & \left(\begin{array}{ccc}
e^{-\kappa t/2} & 0 & 0\\
0 & \frac{e^{-(\kappa+\gamma)t/4}}{F_{{\rm R}}}C_{+}^{{\rm R}} & -\frac{4i\tilde{G}_{{\rm R}}}{F_{R}}\sinh\left(\frac{tF_{{\rm R}}}{4}\right)\\
0 & -\frac{4i\tilde{G}_{{\rm R}}}{F_{{\rm R}}}\sinh\left(\frac{tF_{{\rm R}}}{4}\right) & \frac{e^{-(\kappa+\gamma)t/4}}{F_{{\rm R}}}C_{-}^{{\rm R}}
\end{array}\right),
\end{eqnarray*}
with
\begin{eqnarray*}
F_{{\rm W}}=\sqrt{(\kappa-\gamma)^{2}+16\tilde{G}_{{\rm W}}^{2}}, &  & \,\,\,F_{{\rm R}}=\sqrt{(\kappa-\gamma)^{2}-16\tilde{G}_{{\rm R}}^{2}},\\
C_{\pm}^{{\rm W}}=F_{{\rm W}}\cosh\left(\frac{F_{{\rm W}}t}{4}\right)\pm(\kappa-\gamma)\sinh\left(\frac{F_{{\rm W}}t}{4}\right), &  & \,\,\,C_{\pm}^{{\rm R}}=F_{{\rm R}}\cosh\left(\frac{tF_{{\rm R}}}{4}\right)\pm(\kappa-\gamma)\sinh\left(\frac{tF_{{\rm R}}}{4}\right).
\end{eqnarray*}

\end{widetext}

One immediate application of this formalism is to study cooling in this simplified linearized regime. For an initial thermal state with $n_{{\rm th}}$ mean number of magnons 
\begin{eqnarray*}
\rho(0) & = & \vert0\rangle\langle0\vert_{1}\otimes\vert0\rangle\langle0\vert_{2}\otimes\rho_{{\rm Th,m}},\\
\rho_{{\rm Th,\,m}} & = & \frac{1}{1+n_{{\rm Th}}}\sum_{n\ge0}\left[\frac{n_{{\rm Th}}}{1+n_{{\rm Th}}}\right]^{n}\vert n\rangle\langle n\vert,
\end{eqnarray*}
the temporal evolution under the read Hamiltonian will transform this state into another thermal state. The mean number of magnons $\langle\hat{m}^{\dagger}\hat{m}\rangle$ has temporal evolution given by
\begin{eqnarray*}
\langle\hat{m}^{\dagger}\hat{m}\rangle(t) & = & \left(U_{i2}^{{\rm Read}}(t)\right)^{*}U_{2j}^{{\rm Read}}(t)\langle\hat{A}_{i}^{\dagger}(0)\hat{A}_{j}(0)\rangle\\
 & + & {\displaystyle \int}_{0}^{\,t}d\tau_{1}d\tau_{2}(\left(U_{i2}^{{\rm Read}}(t-\tau_{1})\right)^{*}U_{2j}^{{\rm Read}}(t-\tau_{2})\\
 &  & \times\langle\hat{N}_{i}^{\dagger}(\tau_{1})\hat{N}_{j}(\tau_{2})\rangle)\,.
\end{eqnarray*}
Using the expectation values for the initial state and for the noise operators given in Eqs. \eqref{eq:NoisePhoton} and \eqref{eq:NoiseMagnon},
\begin{eqnarray*}
\langle\hat{m}^{\dagger}\hat{m}\rangle(t) & = & \vert U_{22}^{{\rm Read}}(t)\vert^{2}n_{{\rm Th}}+\vert U_{12}^{{\rm Read}}(t)\vert^{2}\\
 & + & \int_{0}^{t}d\tau\left[\vert U_{12}^{{\rm Read}}(t-\tau)\vert^{2}+\vert U_{22}^{{\rm Read}}(t-\tau)\vert^{2}n_{{\rm Th}}\right].
\end{eqnarray*}
In the limit $t\rightarrow\infty$ this gives the steady state value (for $\tilde{C}_{{\rm R}}=\tilde{G}_{{\rm R}}^{2}/\kappa\gamma$)
\[
\langle\hat{m}^{\dagger}\hat{m}\rangle(t\rightarrow\infty)=\frac{\gamma n_{{\rm Th}}}{(\kappa+\gamma)}\left(1+\frac{\kappa}{\gamma(1+4\tilde{C}_{{\rm R}})}\right),
\]
which is the formula for $n_{0}$ presented in Eq.~\eqref{eq:CoolingFormula} in the main text. For $\gamma\ll\kappa$ and for $\tilde{G}_{{\rm R}}/\kappa\ll1$ this expression is equivalent to the one derived in Ref. \citep{sharmaOpticalCoolingMagnons2018a}, not taking into account a possible heating channel. On the other hand, for strong coupling $4\tilde{C}_{R}\gg1$, the above expression is equivalent to the one used in Ref. \citep{gallandHeraldedSinglePhononPreparation2014} and derived in Ref. \citep{liuDynamicDissipativeCooling2013}, not taking into account quantum backaction.

\section*{Appendix C - Dependence of the heralding probability on the magnon
temperature}

The weak dependence of the heralding probability on the temperature is a consequence of the interplay between the optomagnonic coupling and the thermalization process. For instance, the heralding probability given by Eq.~\eqref{Prob}, depends quadratically on the magnon bath temperature through $1-n_{{\rm Th}}$. The coefficient of $1-n_{{\rm Th}}$ is given by
\[
A_{{\rm Th}}=\frac{8\gamma\tilde{G}_{{\rm W}}\mathcal{F}(t)}{F_{{\rm W}}(\kappa+\gamma)[(\kappa+\gamma)^{2}-F_{{\rm W}}^{2}]},
\]
while the other term, which is $\propto1+n_{0}$, has coefficient
\[
A_{0}=\frac{8\tilde{G}_{{\rm W}}e^{-\frac{\kappa+\gamma}{2}t}}{F_{{\rm W}}}\sinh^{2}\left(\frac{t}{4}F_{{\rm W}}\right).
\]
Asides from the multiplicative factor $8\tilde{G}_{{\rm W}}/F_{{\rm W}}$ common to both $A_{0}$ and $A_{{\rm Th}}$, $A_{0}$ is governed by the timescale $\tau=\left[\frac{(\kappa+\gamma)\pm F_{{\rm W}}}{2}\right]^{-1}$, whereas $A_{{\rm Th}}$ contains, through $\mathcal{F}(t)$, terms proportional to $F_{{\rm W}}$ and $F_{{\rm W}}^{2}$ which in turn depend on the enhanced coupling $\tilde{G}_{{\rm W}}$. Since we are working in a regime in which $\tilde{G}_{{\rm W}}/\kappa$ is small and $\tilde{C}_{{\rm W}}<1/4$, the non-trivial dependence of $A_{{\rm Th}}$ on the coupling is weak and, for  $t<\tau$, $A_{{\rm 0}}>A_{{\rm Th}}$. As the considered timescale increases, the contribution of $A_{{\rm Th}}$ to the heralding probability also increases, and for $t\gg\tau$ 
\[
A_{{\rm Th}}\rightarrow\frac{4\gamma\tilde{G}_{{\rm W}}^{2}}{(\kappa+\gamma)(\kappa\gamma-4\tilde{G}_{{\rm W}}^{2})}
\]
while $A_{0}\rightarrow0$. We emphasize that Eq.~\eqref{Prob} is approximate, and therefore fails in some limits. For instance for $C_{{\rm W}}=1/4$ it exhibits an indeterminacy which is eliminated by considering all higher order terms. Also for stronger couplings, high-order terms are more relevant and need to be taken into account.

\section*{Appendix D - Different linewidth regimes}

The optomagnonic coupling depends on the overlap between the magnon mode function $\bm{m}(\bm{r})$ and the electric-field mode $\bm{E}(\bm{r})$ [see Eq. \ref{eq:CouplingIntegral}]. For spherical YIG samples used in current optomagnonic experiments, this overlap is small. As pointed out in the main text, coherent magnon-photon oscillations can be driven by the read Hamiltonian in the strong coupling limit. One way to achieve such regime is by designing structures that would optimize the coupling. Therefore, in next generation experiments $\gamma/\kappa$ can be different than the value used in the main text.

\begin{figure}
\centering{}\includegraphics[width=1\columnwidth]{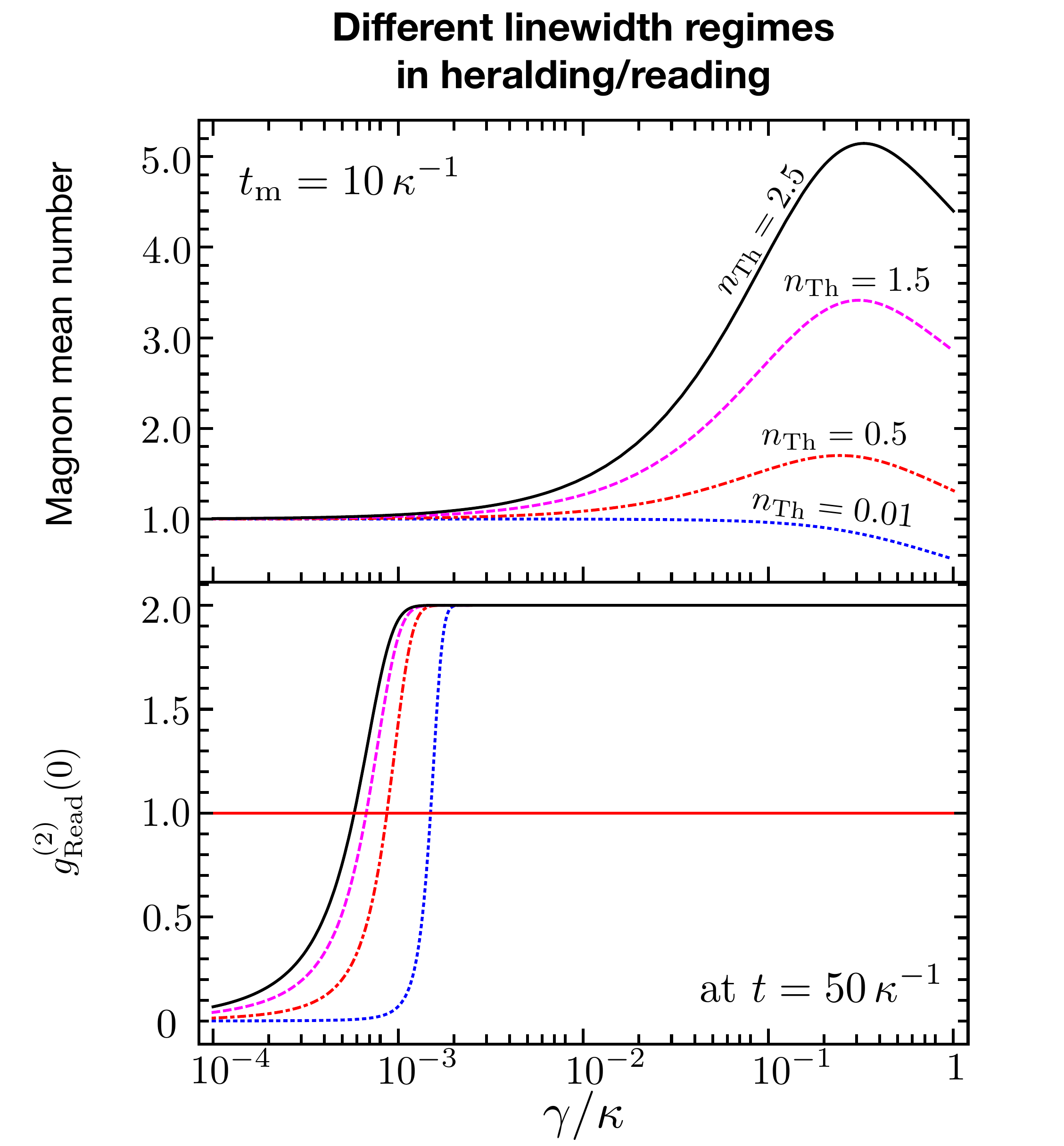}
\caption{Influence of $\gamma/\kappa$ on the heralding protocol. The mean magnon number after the measurement of a photon (upper plot) does not depend on the magnon linewidth for $\gamma\lesssim10^{-3}\kappa$. In this case, thermal effects on the magnon mode can be disregarded. The second-order correlation function for the read mode (bottom plot for $t=50\kappa^{-1}\sim0.5\mu s$ after the beginning of the read pulse) is robust to magnon thermalization for smaller bath temperatures, in accordance with the discussion presented in Fig.~\ref{SecOrdFirst}. For these plots we fixed the cooperativity $\tilde{C}=\tilde{C}_{{\rm W}}=\tilde{C}_{{\rm R}}=10^{-2}$, the time of measurement of a write photon $t_{{\rm m}}=10\kappa^{-1}\sim10^{-7}s$, and the duration of the off phase $T_{{\rm off}}=10^{-2}\kappa^{-1}\sim10^{-10}s$. }
\label{GammaEff}
\end{figure}

In order to analyze the impact of different ratios $\gamma/\kappa$ on the heralding protocol we plot in Fig.~\ref{GammaEff} the mean number of magnons after the measurement of a write photon (upper plot) for different bath temperatures and at a fixed measurement time $t_{{\rm m}}=10\kappa^{-1}\sim10^{-7}s$, together with the second order correlation function of the read mode (bottom plot) for $t=50\kappa^{-1}\sim0.5\mu s$ after the beginning of the read phase. These time scales are in correspondence with those used in Figs.~\ref{MeasProb} and \ref{FigureInitState}. The cooperativity is also fixed to $\tilde{C}=\tilde{C}_{{\rm W}}=\tilde{C}_{{\rm R}}=10^{-2}$. From the plots we see that magnon decoherence is negligible for the heralded state for $\gamma\lesssim10^{-3}\kappa$. In this case the influence of the magnon thermal bath on the system's dynamics is negligible for the time scales considered. On the other hand, for $\gamma/\kappa>10^{-1}$ there is a maximum in the mean magnon number, associated to larger environment influences. The read mode correlation function is bunched for $\gamma\gtrsim10^{-3}\kappa$. The smaller the temperature of the magnon bath, the more robust is the antibunched character of the correlation function with respect to the magnon thermalization process, as was also depicted in Fig.~\ref{SecOrdFirst}.

\section*{Appendix V - Fidelity of the heralded state}

A complementary analysis of the impact of the initial state on the heralding protocol can be done by considering the fidelity of the generated state with respect to a single-magnon Fock state. The fidelity between two states $\rho$ and $\sigma$ is given by
\[
F[\rho,\sigma]={\rm Tr}\left[\sqrt{\rho^{1/2}\sigma\rho^{1/2}}\right],
\]
which can be used as a quantifier of the similarity between the two quantum states \citep{nielsenQuantumComputationQuantum2010}. 

In order to calculate the fidelity of the magnon state after the measurement of a photon, we solve the master equation describing the joint dynamics of the magnon-photon system under the write Hamiltonian

\begin{eqnarray}
\frac{d\rho}{dt} & = & \frac{1}{i\hbar}[\rho,\hat{H}_{{\rm W,1}}]+\kappa_{1}\mathcal{D}_{\hat{a}_{1}}[\rho]+\kappa_{2}\mathcal{D}_{\hat{a}_{2}}[\rho]\label{eq:MasterEquation-1}\\
 &  & +\gamma n_{{\rm th}}\mathcal{D}_{\hat{m}^{\dagger}}[\rho]+\gamma(n_{{\rm th}}+1)\mathcal{D}_{\hat{m}}[\rho],\nonumber 
\end{eqnarray}
where $\mathcal{D}_{\hat{J}}[\rho]=2\hat{J}^{\dagger}\rho\hat{J}-\hat{J}^{\dagger}\hat{J}\rho-\rho\hat{J}^{\dagger}\hat{J}$, using the python QuTiP package \citep{johanssonQuTiPOpensourcePython2012,johanssonQuTiPPythonFramework2013}. We consider the time scales and parameters according with the analysis presented in the main text, and the initial state given by Eq.~\eqref{eq:InitState}. After the temporally evolved density matrix $\rho(t)$ is obtained, the measurement of a write photon is described as
\[
\rho(t)\rightarrow\rho_{{\rm AM}}(t)=\frac{\hat{a}_{1}\rho(t)\hat{a}_{1}^{\dagger}}{{\rm tr}\left[\hat{a}_{1}\rho(t)\hat{a}_{1}^{\dagger}\right]},
\]
through which the reduced density matrix of the magnon subspace can be obtained by tracing out the photon spaces $\rho_{{\rm AM}}^{m}={\rm tr}_{1,2}\left[\rho_{{\rm AM}}(t)\right]$. Finally we compute the fidelity between $\rho_{{\rm AM}}^{m}$ and a single-magnon Fock state $\vert1\rangle\langle1\vert$.

\begin{figure}
\centering{}\includegraphics[width=1\columnwidth]{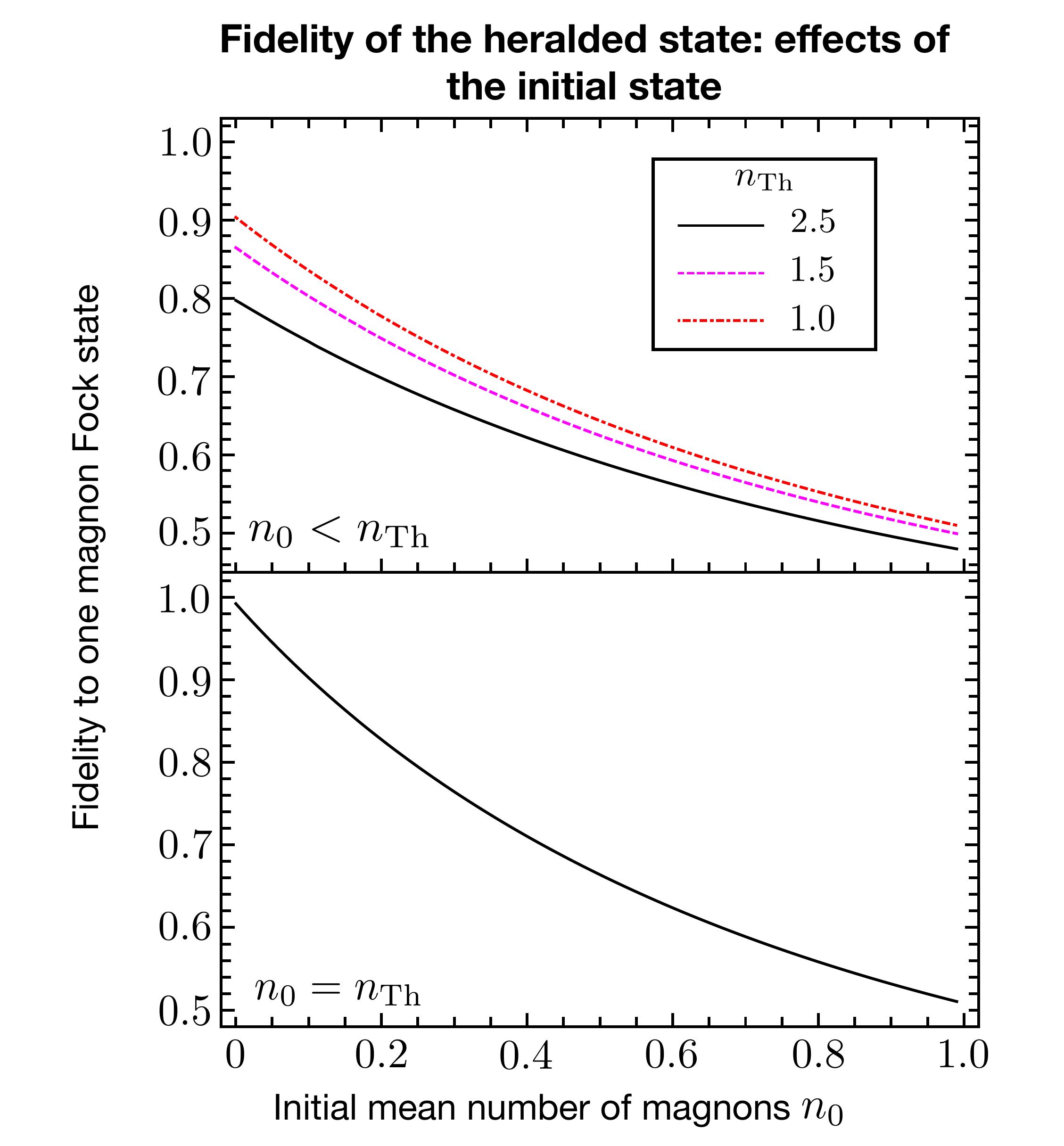}
\caption{Fidelity of the heralded magnon state with respect to a single magnon state as function of the initial number of magnons $n_{0}$ for both an active cooling setup (top plot) and for an initial state in thermal equilibrium with its bath (bottom plot). High fidelities are achieved for small initial number of magnons, prepared through an efficient cooling setup. For this plot we used $\tilde{C}_{{\rm W}}=10^{-2}$, $\gamma/\kappa=10^{-2}$, and $T_{{\rm W}}=t_{{\rm m}}=0.1/\gamma\sim10^{-7}$ s.}
\label{FigureFidelity}
\end{figure}

The results are depicted in Fig.~\ref{FigureFidelity} for an active cooling setup $n_{0}<n_{{\rm Th}}$ (top plot) and for an initial state in equilibrium with the thermal bath $n_{0}=n_{{\rm Th}}$ (bottom plot). The overall dependence of the fidelity with the initial number of magnons is exponential: the more the magnons are in the initial state, the less the heralded state will be closer to a one-magnon Fock state. Good fidelities are obtained for small initial number of magnons, which can be achieved through an efficient cooling setup.

Following this procedure to calculate the fidelity, we obtained the numbers presented in the last paragraph of Sec. V.

\section*{Appendix VI - Gaussian Shaped Pulses}

We can describe the heralding protocol for a setup in which the pulses are not square shaped, but have a more realistic temporal dependence. We adopt a model in which the complete Hamiltonian (in the interaction picture) of the heralding protocol can be written as
\[
\hat{H}_{{\rm Heralding}}=\epsilon_{{\rm W}}(t)\hat{H}_{{\rm W},1}+\epsilon_{{\rm R}}(t)\hat{H}_{{\rm R},2},
\]
where $\epsilon_{{\rm W,R}}(t)$ are functions modeling the temporal shape of the write and read pulses. In particular, we consider in the following Gaussian shaped pulses
\[
\epsilon_{{\rm W,R}}(t)=\frac{1}{\sqrt{2\pi}}\exp\left[-\frac{(t-t_{{\rm W,R}})^{2}}{2\sigma_{{\rm W,R}}}\right],
\]
where $t_{{\rm W,R}}$ are the times corresponding to the maximum intensity of the pumping laser with temporal width $\sigma_{{\rm W,R}}$. The evolution of the system's density matrix is given by the master equation 

\begin{eqnarray}
\frac{d\rho}{dt} & = & \frac{1}{i\hbar}[\rho,\hat{H}_{{\rm Heralding}}]+\kappa_{1}\mathcal{D}_{\hat{a}_{1}}[\rho]+\kappa_{2}\mathcal{D}_{\hat{a}_{2}}[\rho]\label{eq:MasterEquation}\\
 &  & +\gamma n_{{\rm th}}\mathcal{D}_{\hat{m}^{\dagger}}[\rho]+\gamma(n_{{\rm th}}+1)\mathcal{D}_{\hat{m}}[\rho],\nonumber 
\end{eqnarray}
where $\mathcal{D}_{\hat{J}}[\rho]=2\hat{J}^{\dagger}\rho\hat{J}-\hat{J}^{\dagger}\hat{J}\rho-\rho\hat{J}^{\dagger}\hat{J}$,
which we solved numerically with the python package QuTiP \citep{johanssonQuTiPOpensourcePython2012,johanssonQuTiPPythonFramework2013}.

\begin{figure}
\centering{}\includegraphics[width=1\columnwidth]{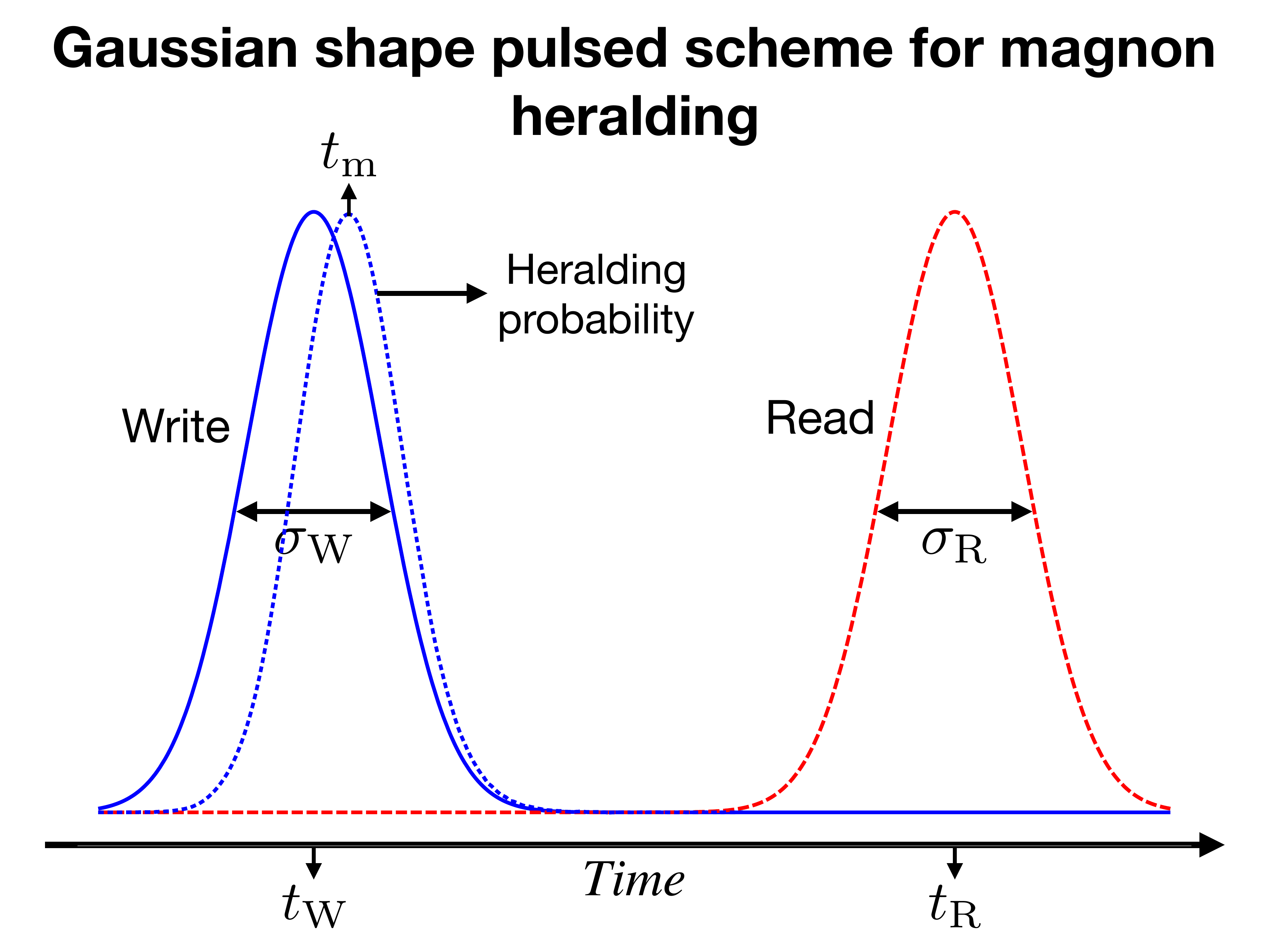}
\caption{Representation of the scheme involving Gaussian pulses. Both write and read interactions are shaped by Gaussian functions centered around $t_{{\rm W}}$ and $t_{{\rm R}}$, respectively, and have widths $\sigma_{{\rm W,R}}$. The dotted Gaussian curve corresponds to the typical behavior of the probability of measuring one write photon as a function of time. We assume that a photon is measured at the instance corresponding to the maximum probability, as indicated in the figure.}
\label{PulsedScheme}
\end{figure}

For this scenario, we assume that a write photon is measured at the time corresponding to the maximum probability of measuring a write photon. In order to compare our results with the ones presented in the main text, we set $t_{{\rm W}}=10^{-7}$ $s$ while the write pulse width is fixed as $\sigma_{{\rm W}}=t_{{\rm W}}$, such that the total area of this Gaussian pulse is similar to the area of the rectangular pulse considered in the main text. The Gaussian pulse scheme is depicted in Fig.~\ref{PulsedScheme}.

\begin{figure}
\centering{}\includegraphics[width=1\columnwidth]{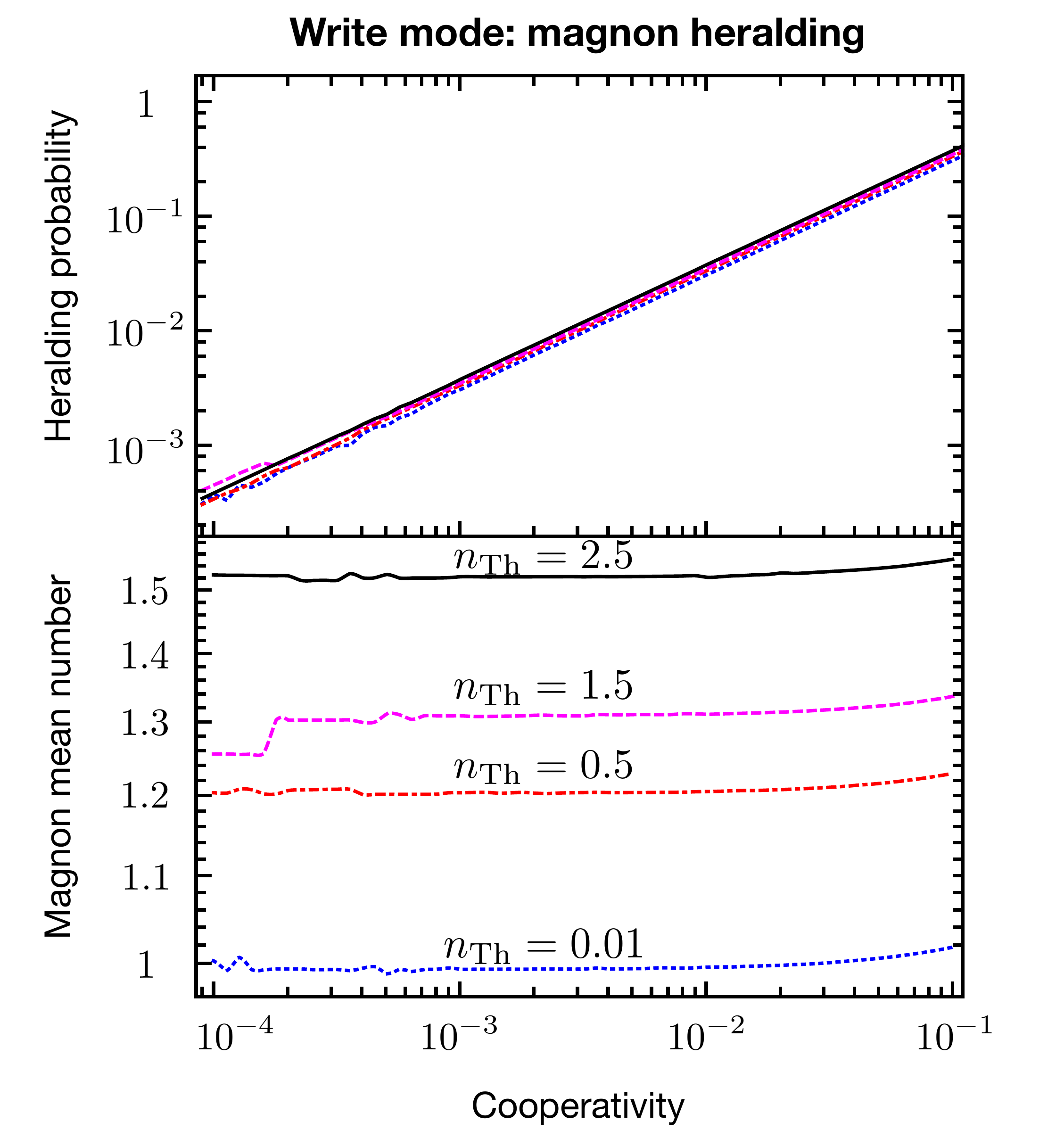}
\caption{Probability of measuring one write photon (top) and mean number of magnons after the measurement of one write photon (bottom) as a function of the write cooperativity for the Gaussian-pulse setup depicted in Fig.~\ref{PulsedScheme}. For this plot the initial state is the vacuum of the system.}
\label{ResultGauss01}
\end{figure}

The results for Gaussian pulses are consistent with the analysis shown in the main text for square pulses. For an initial vacuum state, both the probability of measuring one write photon, and the mean number of magnons after the measurement (see Fig.~\ref{ResultGauss01}), exhibit the same behavior as the results shown in Fig.~\ref{MeasProb}. Nevertheless, the probability of measuring one write photon is smaller than in the square pulse case, and detrimental thermal effects are stronger as it is reflected in the mean number of magnons after the measurement.

\begin{figure}[h]
\centering{}\includegraphics[width=1\columnwidth]{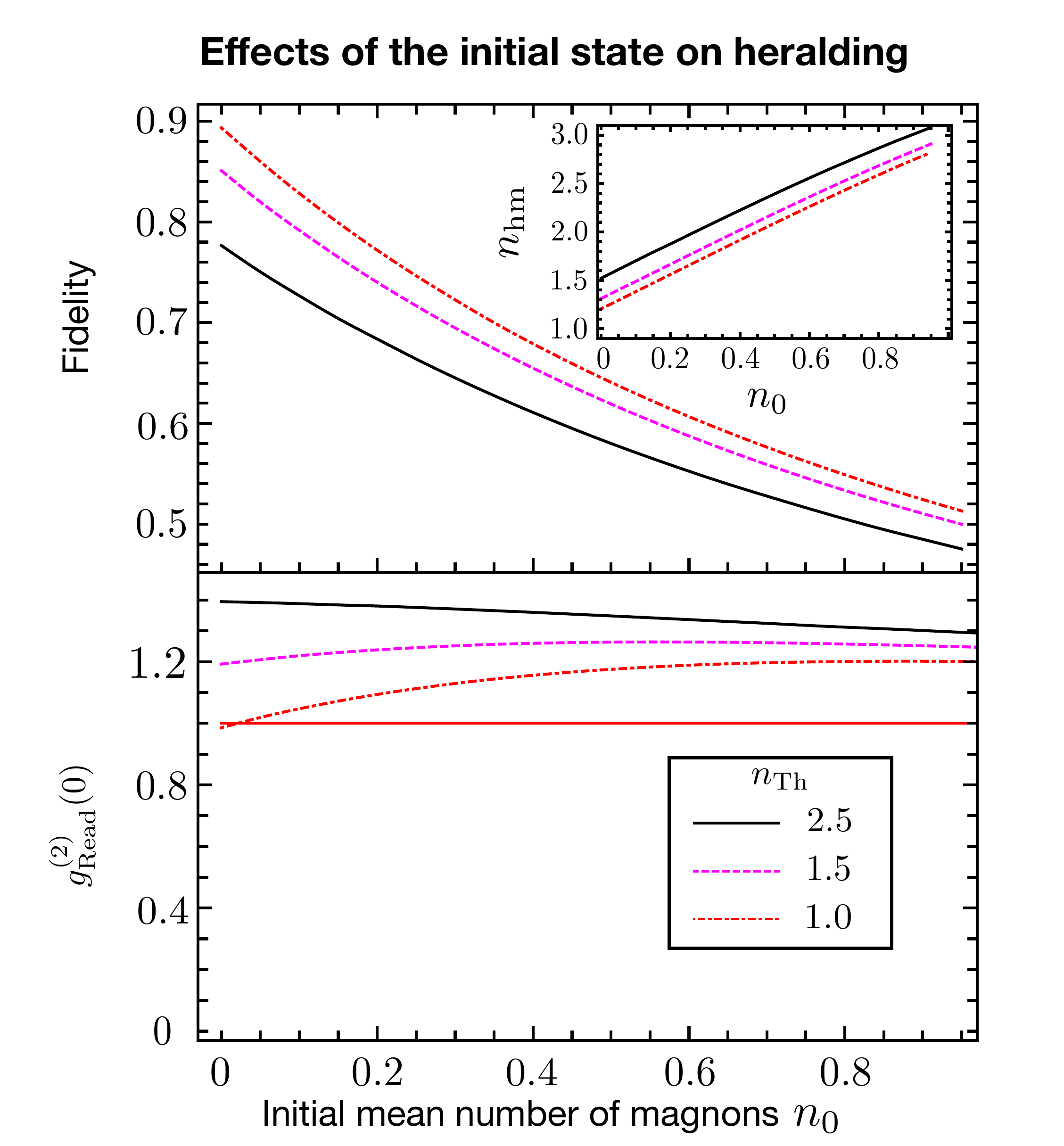}
\caption{Fidelity with respect to a single-magnon Fock state after the measurement of one write photon (top) and second-order correlation function of the read mode (bottom) as a function of the mean number of magnons in the initial state. The detrimental effects are stronger than in the case considered in the main text; the fidelity of the heralded state is smaller than the one obtained with the simplified square pulse analysis (see Fig.~\ref{FigureFidelity}). The second-order correlation function, here calculated at $t=4.3t_{{\rm W}}$ from the beginning of the protocol (according with the time scales of Fig.~\ref{PulsedScheme}), exhibits the same behavior depicted in Fig.~\ref{FigureInitState} but, in this case, the read photon can be bunched even for small $n_{0}$. }
\label{ResultsGauss02}
\end{figure}

The effects of the initial state in this setup are summarized in Fig.~\ref{ResultsGauss02}. The fidelity of the heralded state with respect to a single magnon Fock state decays exponentially with the initial mean number of magnons, as the corresponding case studied in the Appendix C. Here, however, the effects of the initial state are stronger, since the fidelity of the heralded state is smaller compared with the corresponding case in Fig.~\ref{FigureFidelity}. Such detrimental effects are also exhibited by the second-order correlation function of the read photon mode, which becomes bunched faster as a function of the number of magnons in the initial state.

The detrimental effects in the Gaussian-pulse scheme, as depicted in Fig.~\ref{ResultsGauss02}, restrict further the cooling parameters. Apart from cooling the system, such effects can be minimized by reducing the temporal widths of the Gaussian pulses, at the expense of reducing the probability of heralding the state. In general lines the analysis presented in the text agrees qualitatively well with the numerical study of this appendix.

Although presented here as an illustration of more realistic laser pulses, tailoring the temporal dependence of the photon pulses can be used, for instance, to prepare nonclassical states by heralding in a setup in which the shape of the pulses \textquotedbl print\textquotedbl{} the heralded state in a very robust and versatile way \citep{davisPaintingNonclassicalStates2018}.

\FloatBarrier

\bibliographystyle{apsrev}
\bibliography{PRA_Proofs_corrections}

\end{document}